\documentclass[onecolumn,floats,showpacs,prd,amssymb,floatfix,nofootinbib,balancelastpage,superscriptaddress,amsmath,showkeys]{revtex4}
\usepackage{epsfig}
\usepackage{latexsym}
\usepackage{graphicx}
\usepackage{amssymb}

\def\be{\begin{equation}}
\def\ee{\end{equation}}
\newcommand{\ds}{{\sffamily DarkSUSY}}
\newcommand{\Dpp}{D_{pp}}
\newcommand{\Dxx}{D_{xx}}
\newcommand{\ddp}{{\partial\over\partial p}}
\def\mc{\mathcal}

\begin{document}

\title{The contribution to the antimatter flux from individual dark matter substructures}

\author{Marco Regis}
\email{Marco.Regis@uct.ac.za} 

\affiliation{Cosmology and Gravity Group, Department of Mathematics and Applied
Mathematics, University of Cape Town, Rondebosch 7701, South Africa}

\author{Piero Ullio}
\email{ullio@sissa.it}

\affiliation{SISSA,
              Via Beirut 2-4, I-34014 Trieste, Italy and\\
              Istituto Nazionale di Fisica Nucleare,
              Sezione di Trieste, I-34014 Trieste, Italy}

%------------------------------------------------------------------------------

\begin{abstract}

The local antimatter fluxes induced by an individual dark matter (DM) substructure 
can be significantly dependent on the proper motion of the source.
We derive analytic solutions to the propagation equation for time-dependent 
positron and antiproton primary sources, finding that the static limit is a 
fair approximation only for very high energy particles and nearby sources.
We discuss weakly interacting massive particle (WIMP) models fitting 
the PAMELA positron excess and the FERMI all-electron data. We show that, 
for a single non-static DM point-source, one cannot extract from the data, in a unique way,
model independent particle physics observables, such as the WIMP mass, 
the pair annihilation cross section, and the annihilation yield.
The gamma-ray emission associated to WIMP models 
inducing a significant local flux of positrons or antiprotons
is found to be compatible with EGRET measurements, 
but it can be definitely singled out with the FERMI-LAT telescope.

\end{abstract}



%------------------------------------------------------------------------------

\maketitle

\section{Introduction}

The recent measurements of the local cosmic-ray electron and positron fluxes have stimulated
a lot of interest in the cosmic-ray field. The sharp rise in the positron fraction detected by 
PAMELA~\cite{Adriani:2008zr} seems a clear indication of the existence of a nearby positron
source in addition to the secondary positron component due to the interaction of primary species 
with the interstellar medium along propagation. The very recent high-statistic measurement 
by FERMI~\cite{Abdo:2009zk} of the all-electron flux (namely the flux of electrons plus 
positrons without charge discrimination), while not confirming previous hints of an anomalous 
peak by ATIC~\cite{:2008zzr} and PPB-BETS~\cite{Torii:2008xu}, has found a spectral index sensibly 
harder than the one inferred from past extrapolations for the electron spectrum based on lower 
energy (and less accurate) data. Such a picture leaves room for a substantial contribution from unconventional 
(or previously disregarded) lepton sources, and it is very tempting to consider the
possibility that a large component of the positron and electron flux is provided by the pair 
annihilation of dark matter (DM) particles, possibly forming the DM halo of the Milky Way 
(for early proposals in this direction see, e.g.,~\cite{Rudaz:1987ry,Ellis:1988qp,Kamionkowski:1990ty}; 
for reviews on indirect DM detection see, 
e.g., ~\cite{Jungman:1995df,Bergstrom:2000pn,Bertone:2004pz}). The latest fits of the data with 
such a component, see, e.g.,~\cite{Bergstrom:2009fa,Meade:2009iu} confirm that this picture is viable. 
Two general features of the DM-induced source
were well-known even from early analyses: i) to match the level of 
the positron background, the local pair annihilation rate needs to be much larger than the one expected 
for thermal DM weakly interacting massive particles (WIMPs), within standard assumptions for 
the Universe thermal history and the local DM distribution; 
ii) having normalized the annihilation rate to the observed positron flux, there are very stringent 
bounds on the WIMP model from measurements of the local antiproton flux, disfavouring
annihilation modes giving rise to a hadronic yield, and favouring the leptonic channels.
DM models fulfilling these requirements have been proposed recently, see, 
e.g.,~\cite{ArkaniHamed:2008qn,Nomura:2008ru}. For what concerns the first requirement, they
mainly focus on a mechanism to account 
for a mismatch between the thermally averaged annihilation cross section at the freeze-out in the 
early Universe and a much larger annihilation cross section in the halo today. Alternatively,  one could
invoke an enhancement in the positron signal based on the presence of a local population 
of dense dark matter substructures, with the pair annihilation rate being large because the average 
of the number density of DM pairs is much greater than one half of the square of the mean 
DM number density (i.e.,  in terms of the local DM halo density $\rho$,  
$\langle \rho^2 \rangle \gg \langle \rho \rangle^2$). When considering an average effect within 
many realizations of the Milky Way substructure population as extrapolated from current 
$\Lambda$CDM numerical N-body simulations, the mean enhancements in the local 
cosmic-ray fluxes are typically very modest,  possibly below a factor of 
few~\cite{Brun:2007tn,Lavalle:1900wn}.  Large effects, at the level needed to account for PAMELA 
and FERMI data, have been claimed instead in connection to a (few) single, very dense, nearby  
substructure(s)~\cite{Hooper:2008kv,Bringmann:2009ip,Brun:2009aj,Kuhlen:2009is}; the price to
pay in this case is that one has to refer to a configuration with a very small realization probability 
according to the N-body simulations, or to rely on a subhalo picture which is less standard. 

Both in discussing average effects from a full subhalo population and in tracing the effect of individual
substructures, the approach of the recent  papers and the vast majority of papers in the literature 
has been to ignore the fact that one is dealing with a system which is not static, and the emission 
and propagation of charged particles has been treated in the steady state limit. 
The distribution of substructures in the Galaxy is not rotationally supported; their typical velocity 
can be estimated from the total mass density profile of the Milky Way. Assuming for simplicity 
spherical symmetry for the Galaxy and an isothermal sphere for the subhalo phase-space 
distribution function, the velocity dispersion for such a distribution is simply equal to 
$\sqrt{3/2}$ times the value of the circular velocity~\cite{BT},  i.e., assuming 
250~km~s$^{-1}$ for the local rotational speed~\cite{Reid:2009nj}, about  300~km~s$^{-1}$. 
An object moving at such speed on an orbit perpendicular to the Galactic plane has a time 
of crossing of the diffusive halo region for cosmic-rays, say a cylinder with a 4~kpc half-height, 
of about $10^{15}$~s. Such value is comparable to the typical confinement time for
cosmic rays as estimated in the simplified "leaky-box" propagation models, and to the energy loss 
timescale for electrons and positrons~\cite{G}. This is clearly just a qualitative argument 
(actually not even referring to the most  appropriate quantities, see the discussion below) to 
illustrate that the effect of proper motion of DM substructure can be relevant. Indeed, we will show
that there is the possibility that local antimatter measurements reflect a transient due to
DM annihilations in a subhalo, rather than a source to be modeled in the steady state 
limit.

The paper discusses the main features of the local antimatter fluxes resulting from individual
DM substructures, taking proper motion into account. Analytic solutions for the
propagation equation, as appropriate for positrons and antiprotons/antideuterons, 
are presented taking into account the most relevant terms, referring 
to a simplified description of the diffusion region, the interstellar medium and 
radiation field (these are the same kind of assumptions which are usually implemented 
for DM studies in the steady state limit, as well as, most often, to discuss electron/positron 
fluxes from astrophysical sources). We show results in a few benchmark configurations for 
the propagation parameters and for the orbit of the DM source. We introduce a few sample possible 
interpretations for the PAMELA and FERMI positron/electron data, illustrating how sensitive 
one becomes to different assumptions. In particular, we show that, contrary to the picture extensively 
discussed in recent analyses, it is no longer true that one can extract from 
the data, in a unique way, model independent particle 
physics observables, such as the DM mass, the pair annihilation cross section and the 
annihilation channel. We also consider the gamma-ray signals associated to this scenario and compare with current 
limits as well as with the detection prospects in the upcoming future.

The paper is organized as follows. In Section~\ref{sec:CRprop}, we present the description of the particle propagation in the Galaxy. We discuss point-like DM substructures as sources of positrons and compare to PAMELA and FERMI data in Section~\ref{sec:positronpoint} and \ref{sec:positrondata}, respectively. The contribution to the antiproton flux is presented in Section~\ref{sec:antiproton}. In Section~\ref{sec:extra}, we compute other detectable extra features of the electron/positron flux, such as the associated radiative emission and the dipole anisotropy of the spectrum. Section~\ref{sec:concl} concludes. Details on the solution of the transport equation for positrons and antiprotons are reported in the Appendix.

\section{The cosmic-ray propagation model}
\label{sec:CRprop}

The random walk of charged cosmic-rays in the turbulent and regular component of 
the Galactic magnetic fields is usually treated through an effective description. Defining $n$
the  number density per unit total particle momentum of a given particle species
(i.e., $n(p) dp$ is the number of particles in the momentum interval $(p,p+dp)$), the 
propagation equation can be casted in the general form (see, e.g.~\cite{Strong:2007nh}):
\begin{equation}
\frac{\partial n (\vec r,p,t)}{\partial t} = Q(\vec r, p, t)                                             
   + \vec\nabla \cdot ( \Dxx\vec\nabla n - \vec{v}_c n)
   + \ddp\, p^2 \Dpp \ddp\, \frac{1}{p^2}\, n                  
   - \frac{\partial}{\partial p} \left[\dot{p} \,n 
   - \frac{p}{3} \, (\vec\nabla \cdot \vec{v}_c )n\right]
   - \frac{n}{\tau_f} - \frac{n}{\tau_r}\,.
\label{eq:prop_n}
\end{equation}
Here $Q$ is the source term, including primary, spallation and decay contributions,
$\Dxx$ is the spatial diffusion coefficient, $\vec{v}_c$ is the convection velocity, diffusive 
reacceleration is described as diffusion in momentum space and is determined by the 
coefficient $\Dpp$\ , $\dot{p}\equiv dp/dt$ is the momentum gain or loss rate, $\tau_f$ is 
the time scale for loss by fragmentation, and $\tau_r$ is the time scale for  radioactive
decay. The problem is usually solved for stationary sources and assuming $n$ has
reached equilibrium (i.e. setting the left-hand side of the equation to zero), through a
fully numerical integration of the general model, see, e.g., the Galprop~\cite{Strong:1998pw}
and the Dragon~\cite{Evoli:2008dv} packages, or
implementing (a chain of) semi-analytic solutions valid within a set of simplifying 
assumptions, see, e.g.,~\cite{WLG,Maurin:2001sj}. In our analysis we will
follow the second route, implementing however solutions of the diffusion equations 
valid for positron or antiproton primary sources which are non-stationary.
As commonly done, we will assume a spatially constant diffusion coefficient and
introduce a spatial average for the positron/electron energy loss rate; we will also neglect both 
convection and reacceleration, sketching the relative energy dependent effects 
through an appropriate choice of the momentum scaling of the spatial diffusion 
coefficient.  This very simplified scheme, which is flexible enough to introduce many study 
cases without neglecting any of the physical effects we wish to discuss, can actually be 
sufficient for a fair description of some of the key observables in cosmic-ray physics.
E.g., Ref.~\cite{Moskalenko:2001ya} introduces, for the Galprop numerical package, 
the case of a standard diffusive model with spatial diffusion coefficient of the form:
\be
 D(p) = \beta D_0 \left(\frac{R}{R_0}\right)^\delta\,,
\ee
with $\beta$ being the particle velocity in unit of the speed of light, $R$ being its rigidity, and with the following 
parameter choice: $D_0 = 2.5 \cdot 10^{28}\,{\rm cm^2\,s^{-1}}$, $R_0 = 4\,{\rm GV}$, 
$\delta = 0.6$ for $R>R_0$ and $\delta=0$ for  $R<R_0$ (having neglected $\Dpp$,
from now on we will simply label $\Dxx$ as $D$). The diffusion region is treated 
as a cylinder extending from $+h_h$ to $-h_h$ in the vertical direction, with standard
primary sources in a thin layer around $z=0$, and up to $R_h$ in the radial direction; 
in the example of Ref.~\cite{Moskalenko:2001ya}, $h_h = 4$~kpc and $R_h = 30$~kpc. 
Parameters are tuned to reproduce observational constraints, and in particular the relative abundance 
of secondary to primary components. Indeed, running the public available version of 
Galprop within this setup, we find a fairly good fit to the Boron over Carbon ratio data 
(reduced $\chi^2 =  1.23$ for $R>4$~GV, considering the B/C measurements at high energy 
by ATIC~\cite{Panov:2007fe}, CREAM~\cite{Ahn:2008my}, HEAO3~\cite{Engelmann:1990}, and CRN~\cite{Swordy:1990}, and 
having assumed a spectral index for primary nuclei of 2.35 and 2.1 for, respectively  $R<40$~GV and $R>40$~GV),  
and to the antiproton over proton ratio as recently measured by PAMELA.
We label this parameter choice as "model A", and take it as reference benchmark case.  
A comprehensive discussion of the dependence of our results on the propagation model 
is clearly beyond the scope of this paper; the main trends we will present are essentially insensitive 
to slight readjustments in the parameter space. One of the most relevant parameters for our
discussion is the vertical size of the propagation region $h_h$.  It is well known that 
the local measurements of secondary to primary ratios are mostly sensitive to 
$D_0/h_h$ rather than to each of the two parameters. The degeneracy would be broken 
by local measurements of the so-called "radioactive clocks", namely unstable secondaries, 
such as $^{10}$Be as compared to $^9$Be. Such data are however not very accurate at 
present. We consider two extreme setups: a thin halo model with $h_h = 1$~kpc and a thick 
model with $h_h = 10$~kpc; $D_0$ is rescaled to, respectively,  $0.56$ and $4.6$
in $10^{28}\,{\rm cm^2\,s^{-1}}$ as we find simulating this model with Galprop and refitting 
the B/C ratio (reduced $\chi^2 =  1.10$
and $1.22$; the thin halo model is labeled as "model B", while the thick one as "model C"). 
Within these models the local primary proton and electron fluxes are reproduced as well
(for the electron injection spectrum we take a spectral index of 2.30 at high energy, i.e. above $E=4$~GeV). 
The spectra obtained with Galprop in our reference  cases
will be used to as background estimates in the next sections; the local electron and positron 
flux are computed with the spatially-dependent energy loss terms following from the 
standard templates for the interstellar radiation field and the magnetic field profile as 
implemented in the code. 
When computing primary components from WIMP annihilations we will introduce instead 
the simplification of a spatially-constant energy loss term, referring to an average value valid
in the  local neighborhood. In the energy loss configuration "H", which we assume as standard reference, the synchrotron and inverse Compton energy loss terms are driven, respectively, by 
an average magnetic field in the diffusion region which is about $B = 6\, \mu$G, and a mean 
background starlight density $U = 0.75\,{\rm eV\,cm^{-3}}$~\cite{Porter:2008ve}, on top of the 
cosmic microwave background component. For comparison, we will also consider a template 
in which both quantities are sharply reduced, assuming $B = 1\, \mu$G and 
$U = 0.4\,{\rm eV\,cm^{-3}}$ (we label this energy loss configuration by "L").

\section{Positrons from a dark matter point source}
\label{sec:positronpoint}

\begin{figure}[t]
 \begin{minipage}[htb]{8cm}
   \centering
   \includegraphics[width=7.8cm]{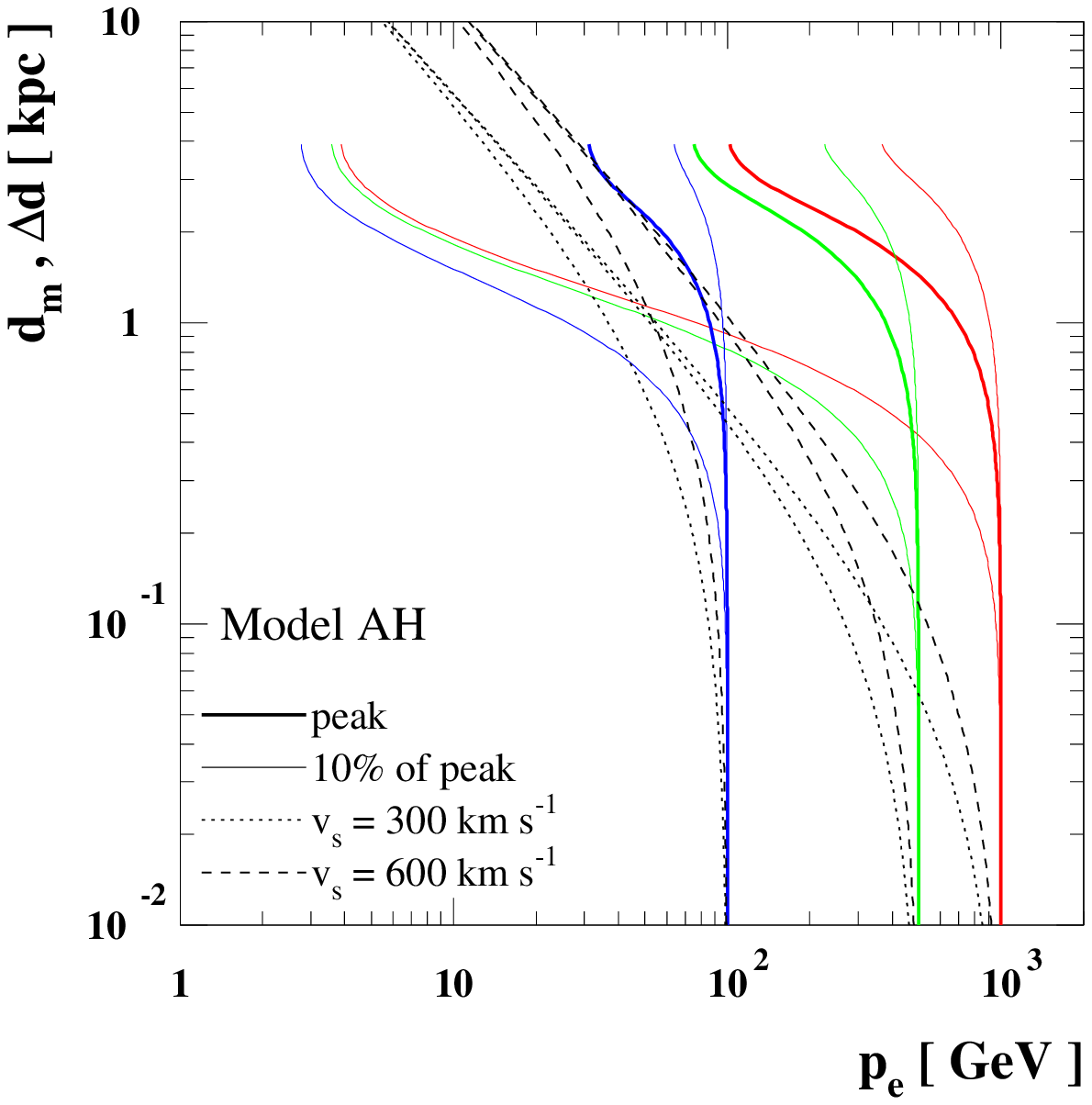}
 \end{minipage}
 \ \hspace{3mm} \
 \begin{minipage}[htb]{8cm}
   \centering
   \includegraphics[width=7.8cm]{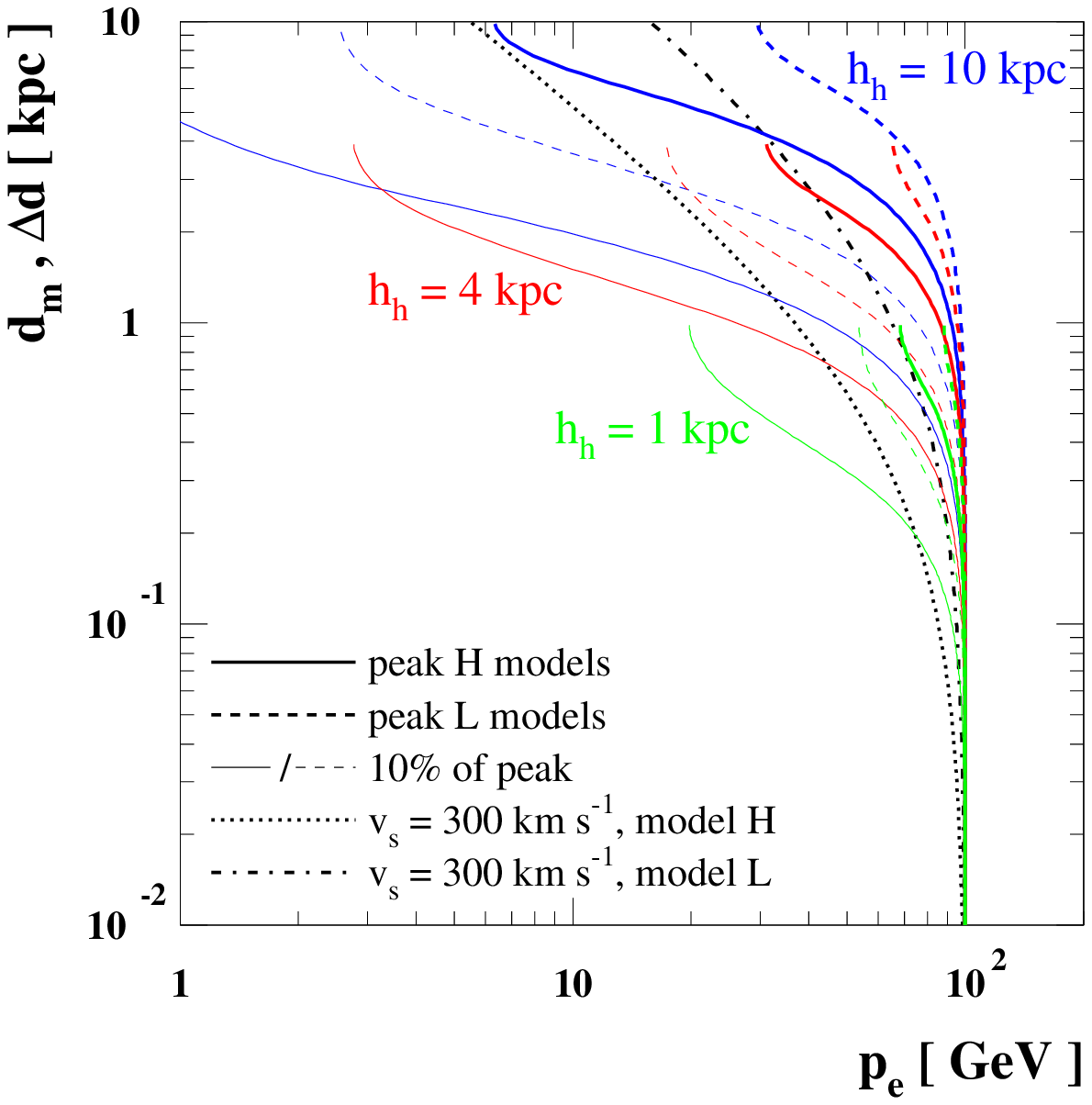}
 \end{minipage}
     \caption{{\it Left Panel:} For three sample values of the positron momentum at injection
     ($p_0 = 100$~GeV, 500~GeV and 1~TeV) and as a function of the positron momentum $p_e$
     as measured locally, the proper motion scale $\Delta d$ (black dashed curves) is compared
     to distance $d_m$ at which the positron Green function is maximal (thick solid curves)
     and the range of values for which the Green function is 10\% of the maximum (thin solid
     curves). The propagation model AH is assumed, see the text for more details.
     {\it Right Panel:} The same as for the left panel, but for $p_0 = 100$~GeV and the six 
     combinations of diffusion and energy loss models.}
\label{fig:epdiff}
\end{figure} 

Consider the limit of a point-like dark matter substructure, entering the diffusion region at the
point $\vec{r}_i$ at the time $t_i$ and moving along an orbit $\vec{r}_p(t)$ (e.g., 
$\vec{r}_p(t) = \vec{r}_i + \vec{v}_s(t-t_i)$ if one can approximate such motion
as a straight  line trajectory with constant velocity $\vec{v}_s$), with dark matter made of 
WIMP dark matter particles of mass $M_\chi$, annihilating in pairs with annihilation rate 
$(\sigma v)$ and positron yield per annihilation $dN_{e^+}/dp$.  The positron dark matter 
source $Q$ at the position $\vec{r}$, momentum $p$ and time $t$, takes the form:
\be
Q(\vec{r}, t,p) = \delta^3 \left[\vec{r}-\vec{r}_p(t)\right]  \frac{dN_{e^+}}{dp} \,\Gamma\,,
\label{eq:pointsource}
\ee
where  $\Gamma$, the total dark matter annihilation rate in the source, contains all terms not 
depending on spatial coordinates, momentum and time:
\be
\Gamma = (\sigma v) \int d\vec{r}_{s} \frac{\rho_{s}^2(\vec{r}_{s})}{2\,M_\chi^2}
\equiv (\sigma v) \frac{\rho_0^2}{2\,M_\chi^2}\,{\mc V}_s
\,.
\ee
Here $\rho_{s}(\vec{r}_{s})$ is the dark matter density profile within the substructure,
and the expression after the equivalence sign defines the annihilation volume ${\mc V}_s$,
having normalized $\rho_{s}$ to the reference value $\rho_0=0.3$~GeV~cm$^{-3}$. 
The positron number density per unit momentum is given by:
\be
n(\vec r,p,t) =  \frac{\Gamma}{\left|\dot{p}(p)\right|}  \int_{t_i}^t dt_0 \int_p^{p_{\rm max}} dp_0\;  
G\left(\vec r,t,p;\vec {r}_p(t_0),t_0,p_0\right) \,  \frac{dN_{e^+}}{dp_0} 
\label{eq:epnsol}
\ee
The Green function $G$ is given in Appendix~\ref{app:pos}; neglecting boundary conditions it is 
approximately in the form:
\be
G \simeq \frac{1}{\pi^{3/2} [\lambda(p,p_0)]^3}
\exp\left\{-\left[\frac{d(t_0)}{\lambda(p,p_0)}\right]^2\right\}  
\delta\left[(t-t_0)- \Delta \tau (p,p_0)\right] \,
\label{eq:epgreensimply}
\ee  
where we introduced the distance $d=\left|  \vec r -\vec r_p(t_0)\right|$ between the source 
and the observer at the time $t_0$, the energy loss timescale  
$\Delta \tau = \int_p^{p_0}  {d\tilde{p}}/{\left|\dot{p}(\tilde{p})\right|}$, and diffusion length $\lambda$,
defined through $\lambda^2 = 4 \int_p^{p_0}  {d\tilde{p}}  \, D(\tilde{p})/{\left|\dot{p}(\tilde{p})\right|}$.
Looking at this expression, we can guess that including the time dependence in the propagation
equation is relevant whenever the variation of $d$ within the time $\Delta \tau$,
say $\Delta d \sim v_s \cdot \Delta \tau$, is larger or comparable to $\lambda$. Consider the high 
energy limit, in which the energy loss term scales like $\dot{p}(p) \propto p^2$ and the diffusion 
coefficient like $D(p) \propto p^\delta$, with $\delta \sim 0.3-0.7$; the square of the ratio 
between $\Delta d$ and $\lambda$ goes like:
\be
\frac{\Delta d^2}{\lambda^2} \simeq  5 \cdot 10^{-2}
\frac{v_{300}^2}{D_1\,\dot{p}_1}
\frac{1-\delta}{1-0.6}
\frac{100^{0.6-\delta}}{p_{100}^{1+\delta}}
\frac{(1-{\mc R})^2}{1-{\mc R}^{1-\delta}}
\label{Eq:ratiodl}
\ee
where $D_1$ and $\dot{p}_1$ are reference values for the diffusion coefficient and energy loss
rate at 1~GeV, respectively, $10^{28}\,{\rm cm^2\,s^{-1}}$ and $-10^{-16}\,{\rm GeV \,s^{-1}}$, 
$v_{300}$ the substructure velocity in units of $300 \,{\rm km \,s^{-1}}$, $p_{100}$ is the positron 
momentum in the equilibrium distribution in units of 100~GeV, and ${\mc R}\equiv p/p_0$ 
the ratio between such momentum and the momentum of the positron at the source.
Being energy losses very effective at high energy, the time spent in the system before losing most of the energy by particles injected with very high momentum tends to be very small. The distance traveled by a DM subhalo in the same timescale is very small as well, and, in this case, proper motion can be safely neglected. An analogous picture occurs when considering cases with small difference between momentum at injection and momentum at equilibrium, and thus with short timescale associated to the electron/positron transport.  
One sees from Eq.~\ref{Eq:ratiodl} that $\Delta d \ll \lambda$ for ${\mc R}\rightarrow 0$ or large $p$.
On the other hand, we expect proper motion to be relevant at intermediate to low energies. 
This effect is illustrated also 
in Fig.~\ref{fig:epdiff}: in the left panel, for a few values of the momentum at the source $p_0$, 
we compute $\lambda$ as a function of $p$ and, correspondingly, find the value of $d_m$, namely,
the distance from us along the direction perpendicular to the Galactic plane at which the 
exact Green function $G$ reaches its maximum (thick solid lines). Neglecting
boundary conditions as in Eq.~(\ref{eq:epgreensimply}), we would simply find 
$d_m = \sqrt{3/2} \,\lambda$; in the plot the scaling of $d_m$ with $p$ gets more rapid
when the distance approaches the vertical boundary of the diffusion region at $h_h=4$~kpc
(the propagation model "AH" is assumed in the plot). Also displayed are the range of distances 
within  which the Green function is larger than one tenth of its maximum and our estimate 
for $\Delta d$, as defined above. In the right panel, we plot the same quantities  for a single 
injected momentum ($p_0=100$~GeV), but looping over the three propagation models and 
the two energy loss configurations as specified in the previous Section. In both panels, 
comparing $d_m$ and the 10\% range to $\Delta d$, one sees that there are large $p$
intervals over which proper motion effects are expected to be large.

\begin{figure}[t]
 \begin{minipage}[htb]{8cm}
   \centering
   \includegraphics[width=7.8cm]{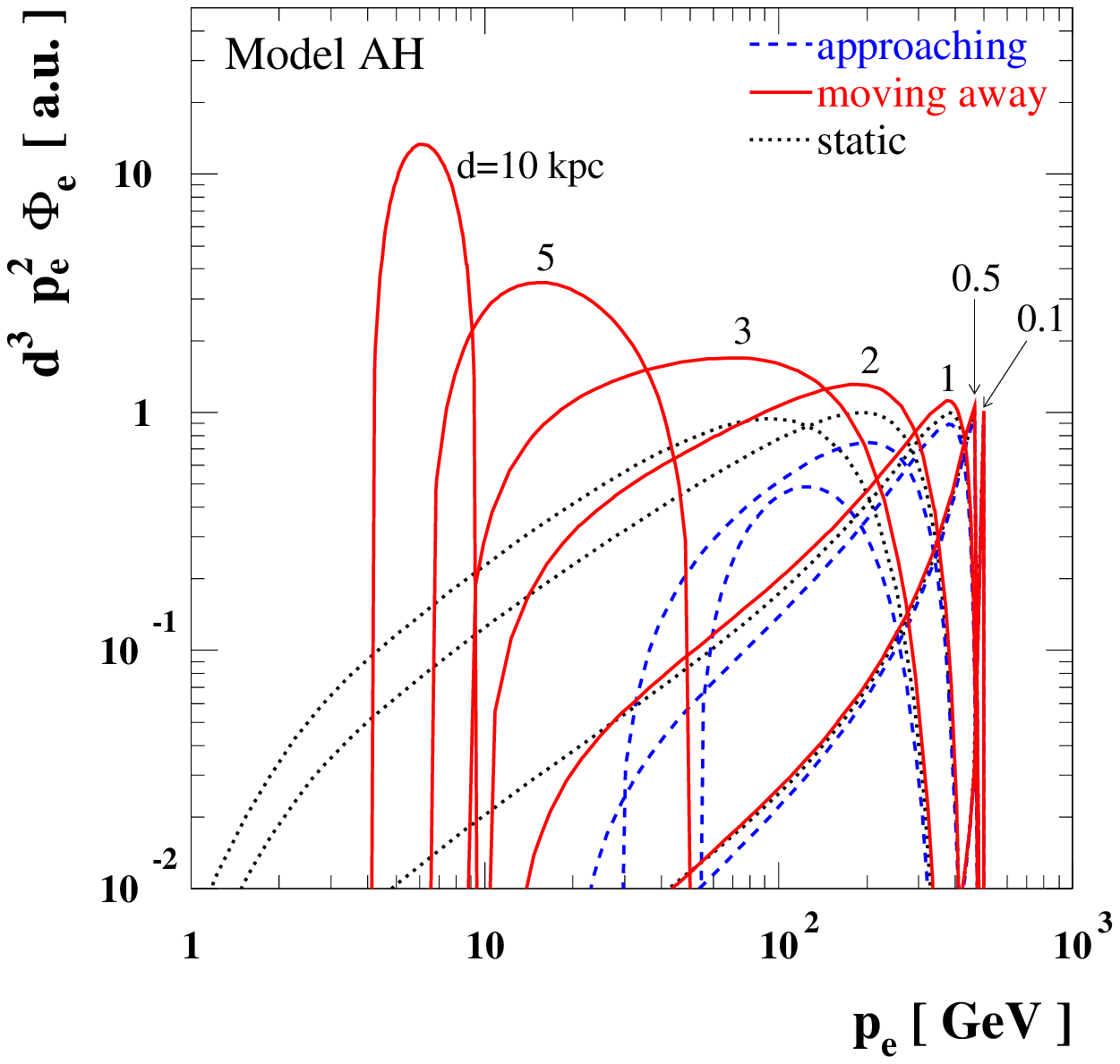}
 \end{minipage}
 \ \hspace{3mm} \
 \begin{minipage}[htb]{8cm}
   \centering
   \includegraphics[width=7.8cm]{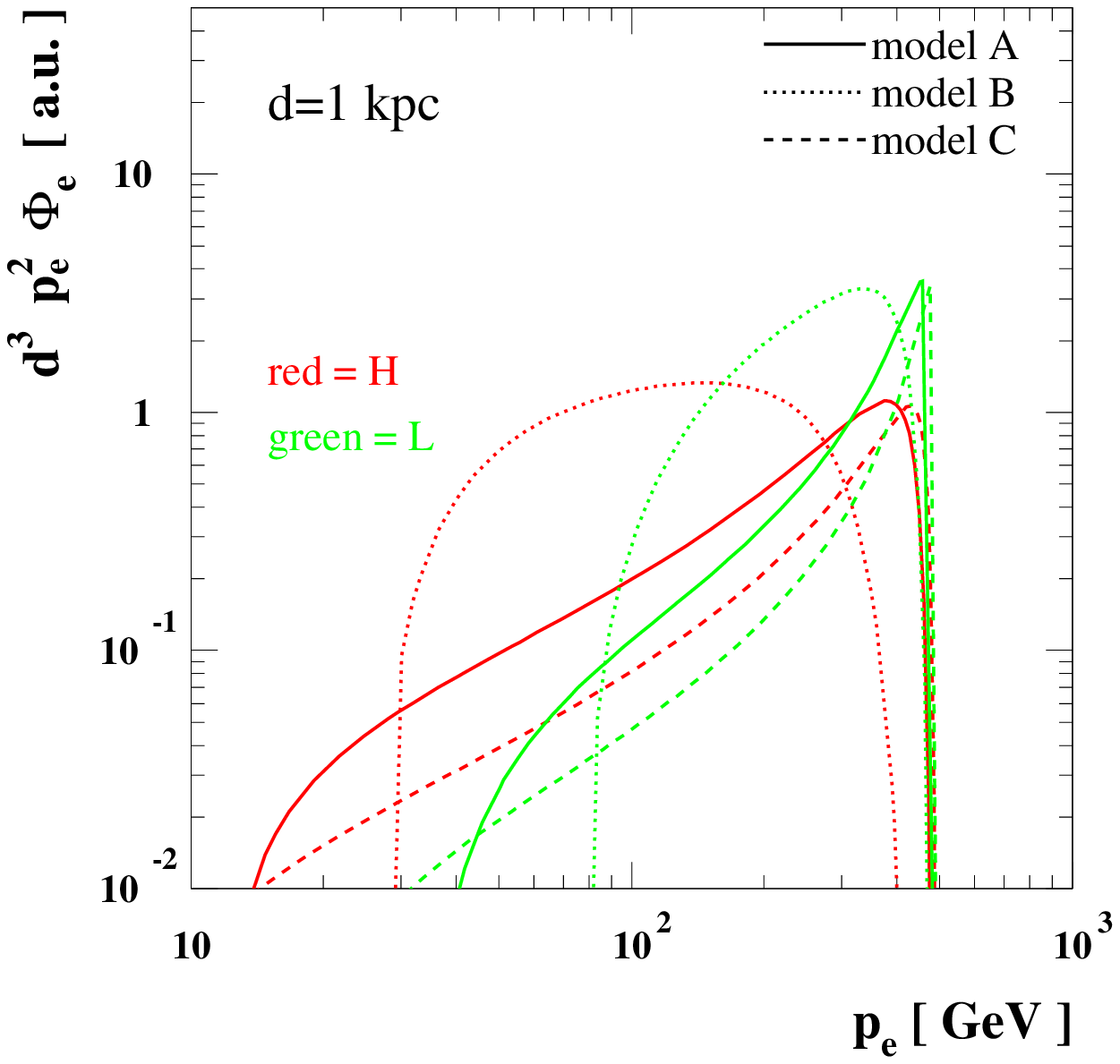}
 \end{minipage}
     \caption{{\it Left Panel:} Local positron flux spectral shape for a point-source composed by 
     500~GeV WIMPs annihilating into monochromatic $e^+\,e^-$, and moving along a vertical 
     orbit intersecting the Galactic plane at a short distance from the observer. 
     Each curve refer to a different time of observation,
     with the distance $d$ of the source from the observer playing the role of time 
     variable. This is useful since we can compare with the spectra one obtains for static 
     sources  placed at the same distances. The positron flux is multiplied by $p^2\,d^3$ to 
     match the approximate scaling in the static limit. {\it Right Panel:} For a give time of
     observation (namely, fixing $d=1$~kpc), predictions for the flux within our benchmark 
     setups for the propagation model.}
\label{fig:scal}
\end{figure} 

To discuss the contribution to the local positron flux $\Phi_e$ from a single point source,
we start considering the sample case of a source moving with constant velocity 
$v_s = 300 \,{\rm km \,s^{-1}}$, on a trajectory which is perpendicular to the Galactic plane, 
intersecting the plane at a small distance from the Sun (say 10~pc, the precise value has 
little relevance for the discussion). To emphasizes propagation effects, we refer to a dark matter 
model, of given mass and total annihilation rate, with prompt annihilation into $e^+\, e^-$ pairs, i.e. 
with a monochromatic positron spectrum $dN_{e^+}/{dp} \simeq \delta (p-M_\chi)$. In 
Fig.~\ref{fig:scal} we fix $M_\chi=500$~GeV and show the shape of the induced local 
positron fluxes in a few sample cases. 
In the left panel, each curve refers to a different time of observation, but rather than labeling
them by time steps, we indicate the distance $d$ of the source from the observer, along the 
trajectory and at the time when the positron flux is measured (dashed lines for an approaching
source, solid line for a source which has passed nearby and now is moving away; for reference, 
in the example we are considering, a shift of 1~kpc in $d$ corresponds to a time interval 
$\Delta t \simeq 10^{14}$~s, i.e. about an order of magnitude lower than the typical 
"escape time" for particles of these energies as extrapolated in simple "leaky box"
propagation models). For comparison, we also show the case of 
static sources (dotted lines), at a given distance from the Sun. 
From the Green function solution (again sufficiently away from the boundaries of the 
propagation region) we can guess that, in the static limit, the peak of the product 
$p^2 \cdot \Phi_e$ scales like $d^{-3}$. This is indeed what we find in the plot, where the
quantity  $d^3\,p^2\,\Phi_e$ has been normalized to 1 in arbitrary units for $d=10$~pc; 
the figure shows also 
the departure from such scaling in case of a non-stationary source, with the result for 
approaching, static and departing sources essentially coinciding only in the limit of small 
distances and large energies.  
For moderate to large distances the height of the peaks increases (decreases) for 
a source which has passed by (is approaching). The spectral shapes are also very different,
with the sharp cutoffs at low energy enforced by the mismatch between the time interval 
from positron emission to detection and the energy loss timescale, which increases at lower
energies. Positron propagation is treated according to model "AH" as introduced above;  such 
model has  $h_h = 4$~kpc and hence an approaching source at $d=3$~kpc, along the vertical
trajectory we are considering, has just entered the diffusion region and induces a local positron 
flux which is marginal compared to the flux from the same source after passing by and being on 
the other side of the trajectory at the same distance. Even after the source has left the diffusion
region there is still a population of positrons which has been left behind contributing to the
local flux, especially at low energies: here a larger $d$ just reflects a longer time interval since 
the injection time and hence a more efficient degrading of the injected high energy positrons 
to small energies. In the right panel of  Fig.~\ref{fig:scal}, we take one of the cases plotted in the
left panel, namely a source moving away from the observer and being at $d=1$~kpc at the 
time of observation, and show instead the dependence of the spectral shape and the normalization 
of the measured flux on the benchmark models chosen for propagation and energy losses. Indeed 
the differences between the various cases are rather striking;  % to some extent they are maximized
this gives a feeling for the fact that there are large uncertainties related to propagation in our 
problem, which are not resolved tuning the model, as we did, to local measurements of 
secondary to primary nuclei ratios. Furthermore, there are potentially other relevant 
issues which we do not address in our simplified approach. E.g., we expect some dependence 
of the results on how boundary conditions are treated in the diffusion model:
we are adopting here the standard approach of a spatially constant diffusion coefficient
and free escape (i.e. the diffusion coefficient going to infinity) as a sharp transition at the vertical
boundaries; this is the same approach followed, e.g., in Galprop. Considering a less sharp
transition between the two regimes, with, e.g., an exponential vertical scaling of the diffusion 
coefficient as proposed by~\cite{Evoli:2008dv},  we would obtain smoother cutoffs in 
the spectra of sources at large distances shown in the left panel of  Fig.~\ref{fig:scal}. A 
gradient in the energy loss term would also mildly affect the prediction for the local positron 
flux.

\begin{figure}[t]
 \begin{minipage}[htb]{8cm}
   \centering
   \includegraphics[width=7.8cm]{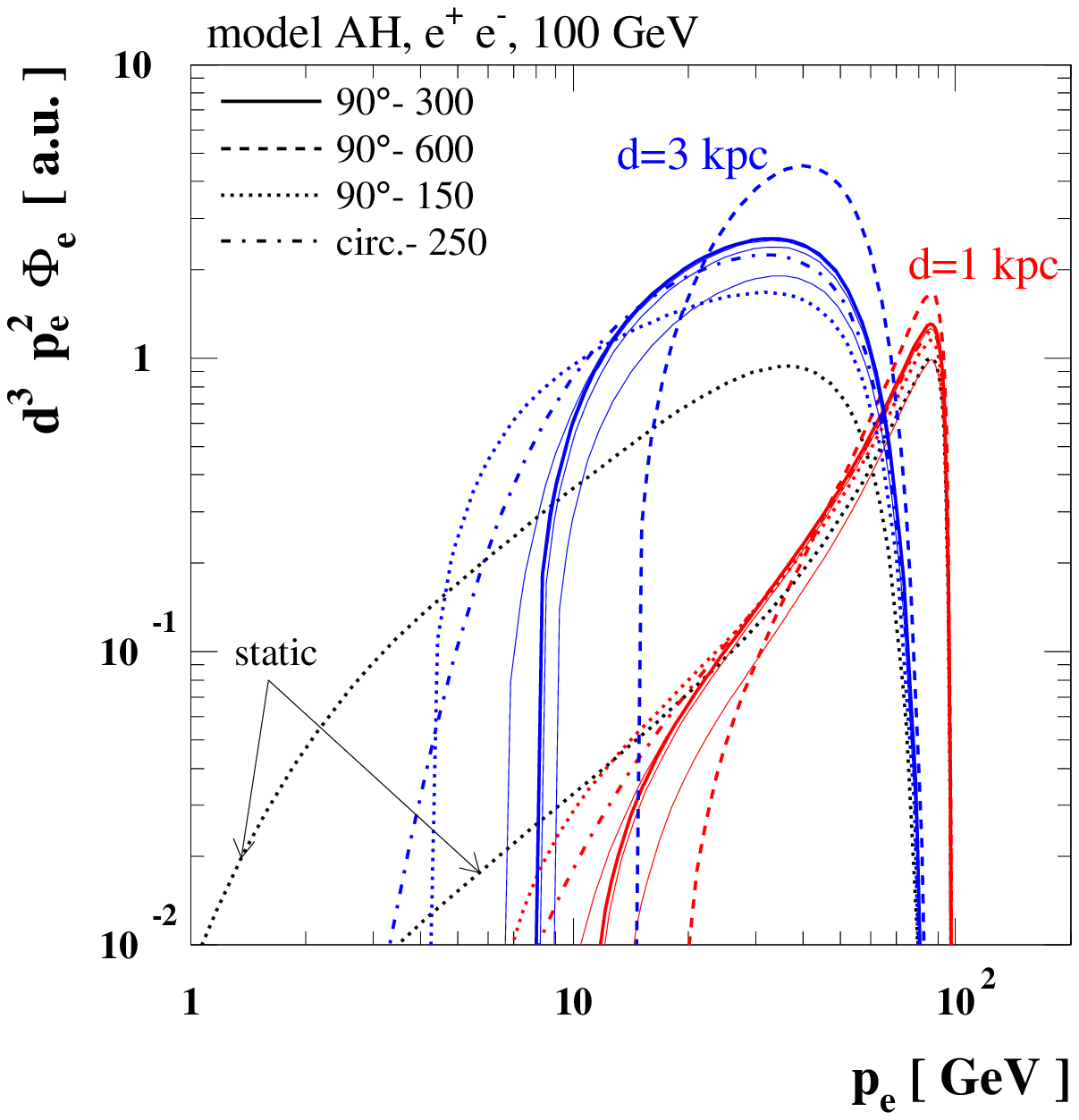}
 \end{minipage}
 \ \hspace{3mm} \
 \begin{minipage}[htb]{8cm}
   \centering
   \includegraphics[width=7.8cm]{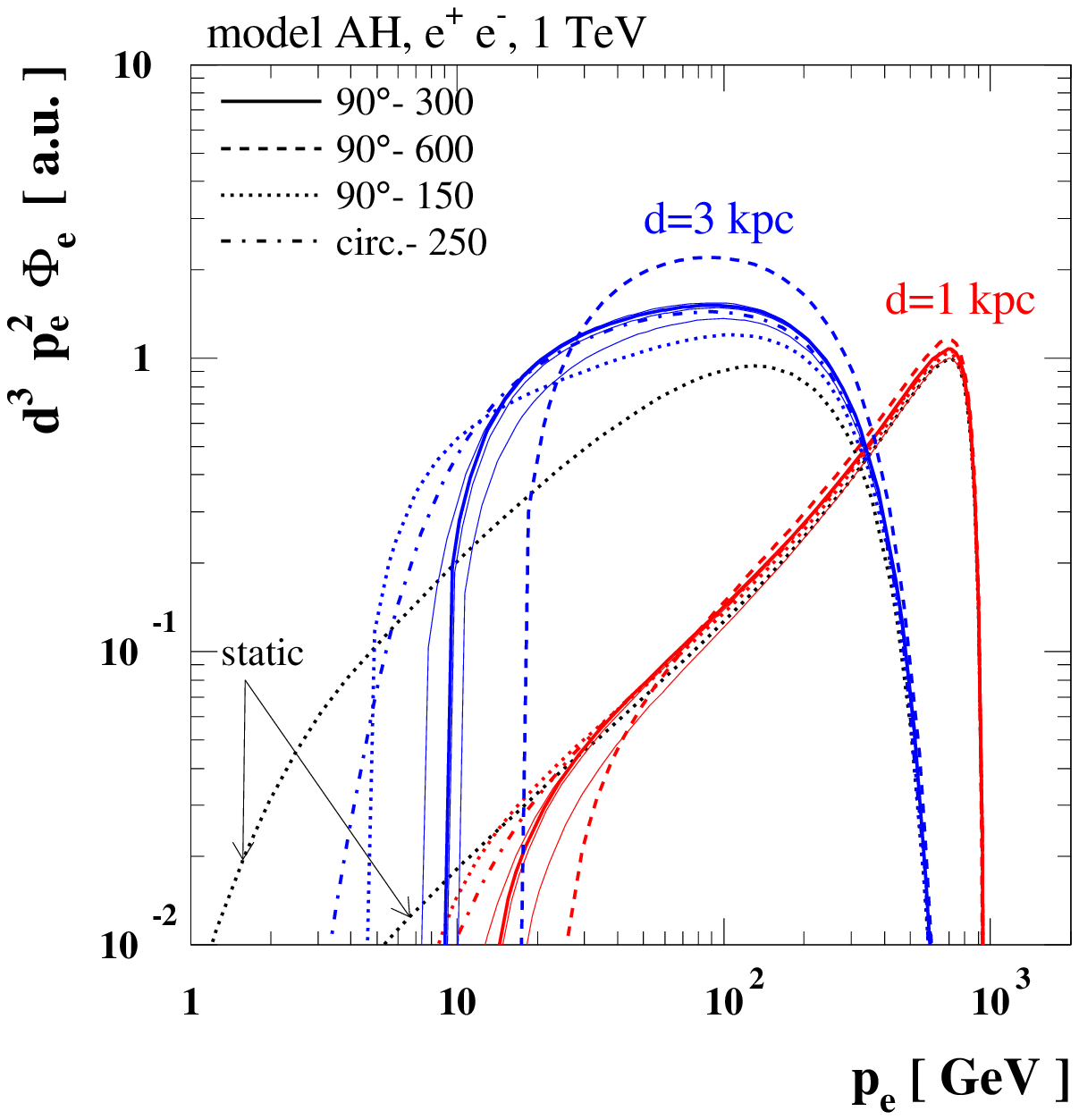}
 \end{minipage}
     \caption{Local positron flux spectral shape for a point source made of 100~GeV 
     (left panel) or 1~TeV (right panel) WIMPs annihilating into monochromatic $e^+\,e^-$, in 
     the benchmark propagation model AH. The two values of $d$ refer to
     the distances which the source has from the observer at the time the flux is measured and
     after the source has passed in the vicinity of the observer; the set of curves displayed refer 
     to a few sample orbits for the source (see the text for details).} 
\label{fig:scal2}
\end{figure}

Fig.~\ref{fig:scal2} considers the cases for a lighter and heavier WIMP (respectively, 100~GeV and 
1~TeV) and, referring again to the propagation model AH,  discusses the dependence of the locally
measured positron flux on the orbit of the source, having chosen two sample values of the present distance of the source from the observer (1~kpc and 3~kpc and source moving away from the 
observer). The thick solid lines correspond to the source moving along a vertical orbit intersecting 
the Galactic plane at 10~pc  from the observer (i.e.,  the same orbit as in 
Fig.~\protect{\ref{fig:scal}}), while thin solid lines are obtained in a few cases in
which this scale is varied up to  1~kpc for $d = 1$~kpc, and to 2~kpc for  $d = 3$~kpc,
or inclining the orbit from vertical to a 45$^\circ$ incident angle; as it can be seen, such cases 
are essentially equivalent and show that what is actually relevant for the result is how much 
time the source spend within a distance from the observer corresponding to diffusion 
length $\lambda$ (with $\lambda$ depending on the emission and observation energies). 
Such time is about the same for all solid lines, while it changes if the source speed is changed.
The dashed and dotted lines refer to the same orbit as for the thick solid line but with a 
source speed which is, respectively, twice as large and half of it, namely 600~km~s$^{-1}$ 
and  150~km~s$^{-1}$ versus 300~km~s$^{-1}$. The dash-dotted lines refer instead to a 
circular orbit with velocity  matching the observed circular velocity, i.e. about  250~km~s$^{-1}$. 
For comparison spectral shapes in the static limit are also displayed. The picture  emphasizes 
further the fact that the encounter with a dark matter point-source determines a transient 
in the local positron flux, and that the static approximation is roughly valid only for nearby sources, 
for which the energy of the measured positrons is not significantly different from the energy at 
emission, or for very energetic positrons. 

We have focussed the discussion on DM sources with a monochromatic positron spectrum, still
it can be very simply extended to DM sources with a generic positron emission spectrum (which
can be obviously thought as a superposition of line spectra for different masses and annihilation 
rates), along the same patterns regarding the dependence on distance and energy.

\section{A non-static point source versus PAMELA and FERMI electron/positron data}
\label{sec:positrondata}

As an application of the general discussion outlined in the previous Section, and to introduce 
a focus on a specific case despite the many ingredients and parameters involved in the problem,
we consider the possibility of a major contribution from a single non-static DM point-source 
to the local electron/positron spectrum as recently measured by PAMELA~\cite{Adriani:2008zr}  and FERMI~\cite{Abdo:2009zk}.  

\begin{figure}[t]
 \begin{minipage}[htb]{8cm}
   \centering
   \includegraphics[width=7.8cm]{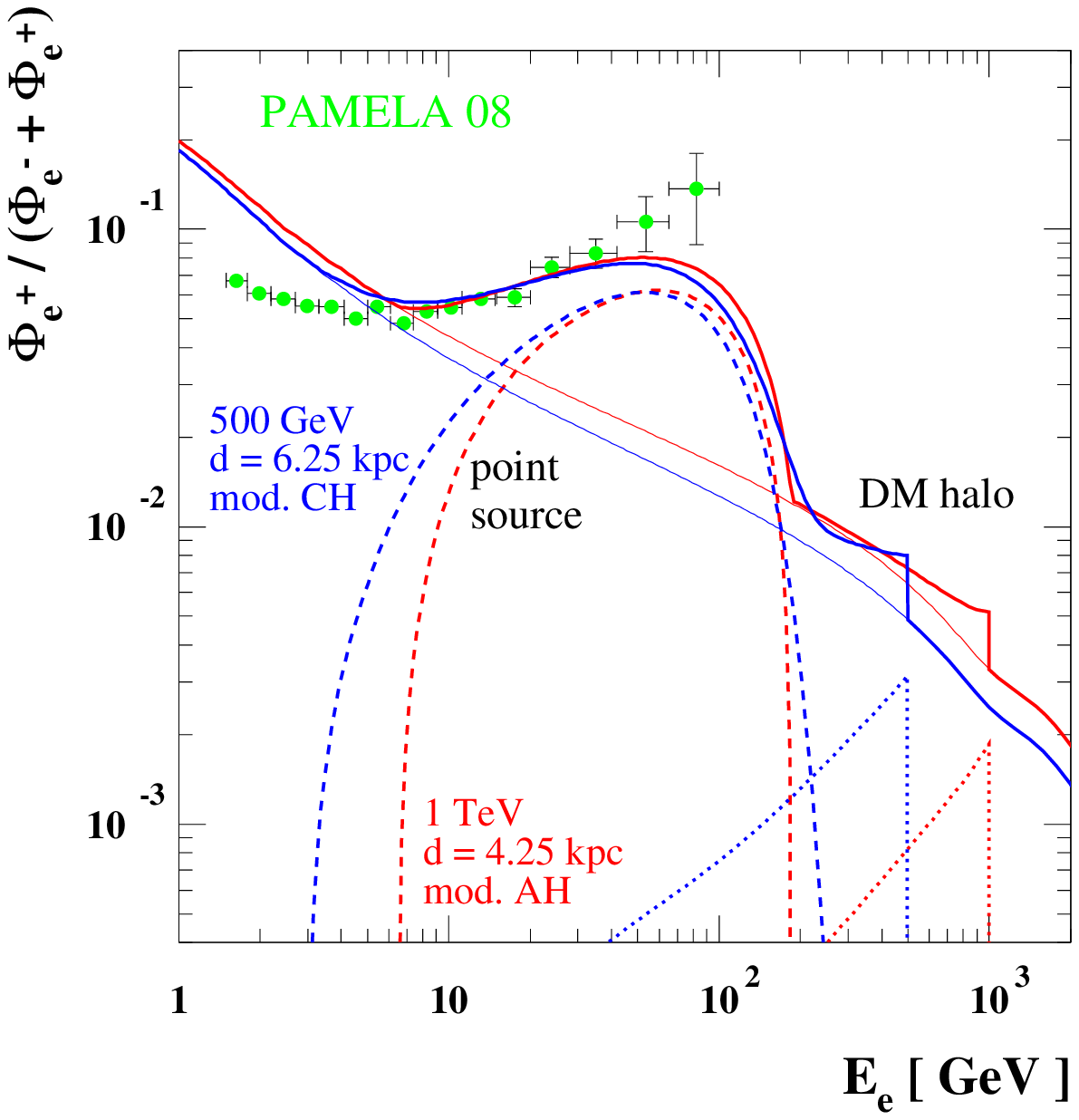}
 \end{minipage}
 \ \hspace{3mm} \
 \begin{minipage}[htb]{8cm}
   \centering
   \includegraphics[width=7.8cm]{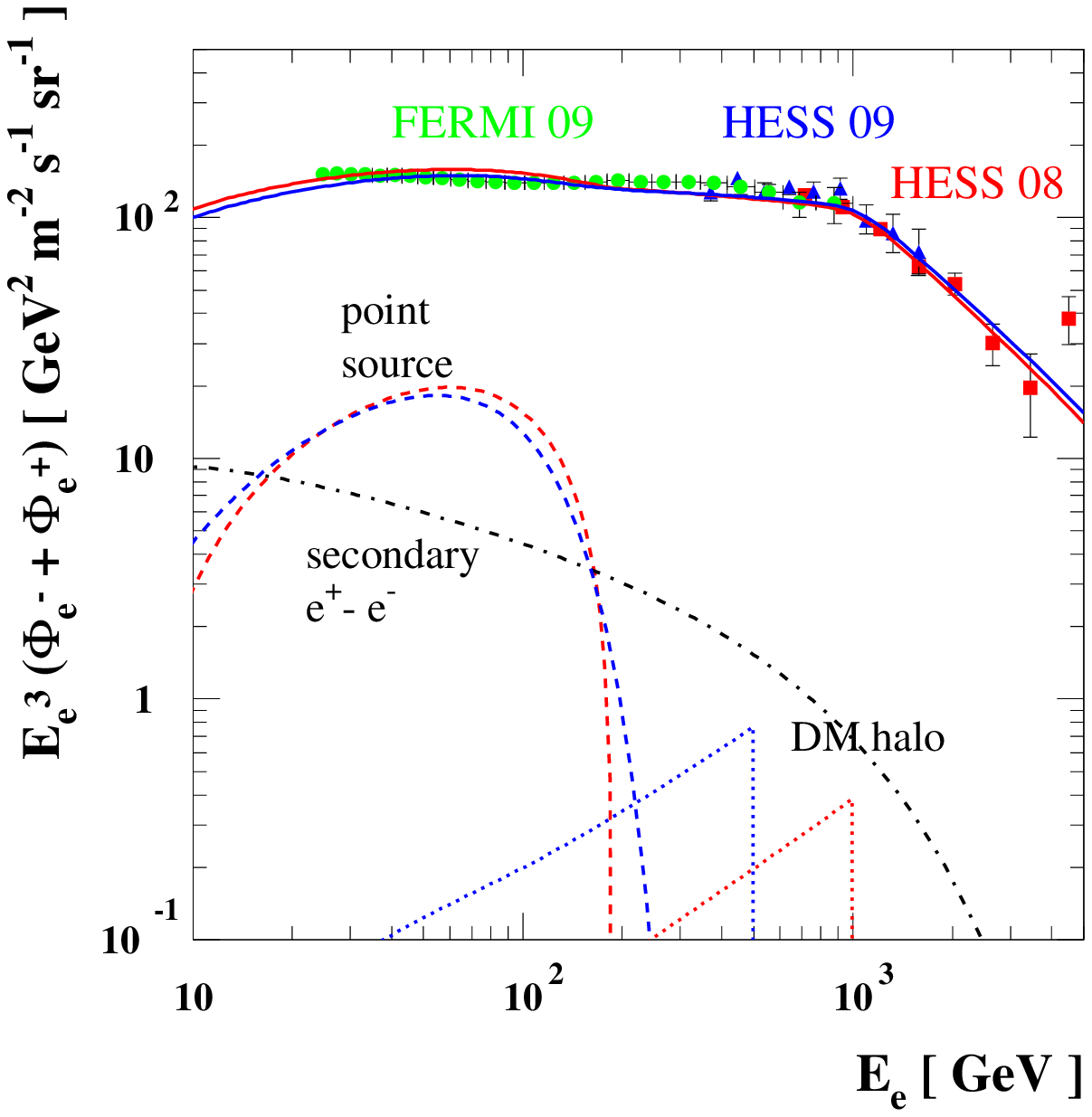}
 \end{minipage}
     \caption{Two examples of fits of the PAMELA positron fraction (left panel) and of the sum of
     the electron and positron fluxes (right panel) with a component due to DM annihilations in a
     substructure. The monochromatic $e^+\, e^-$ final state of annihilation has been considered, 
     as well as two sample values of the DM mass. 
     The fit was performed assuming that the primary electron spectral index 
     and normalization follows from the FERMI data.} 
\label{fig:pamela1}
\end{figure}

For what concerns the background from ordinary astrophysical electrons and positrons,
we will consider two possibilities. 
In the first scenario, the bulk of the "all-electron" spectrum measured by FERMI is due to 
primary electrons emitted in supernova cosmic-ray sources (with only a mild contamination, at the 10-20\% level, from secondary electrons and positrons), and, hence, we can use this data-set to 
derive the electron spectral index at the sources. The second possibility is that the FERMI 
measurement has actually found an extra electron and, possibly, positron source, comparable 
to standard contributions at about 100~GeV or so, and dominating at high energy up to the
cutoff found by HESS at about  1~TeV~\cite{Collaboration:2008aaa}; in this second scenario, we assume that the primary
electron spectrum is softer and can be inferred from preliminary (less accurate) lower energy
data on the electron-only spectrum presented by PAMELA~\cite{PAMELA:preliminary}, 
while the extra contribution is dominated by a single DM source.

Starting with the first background choice, we show in Fig.~\ref{fig:pamela1} (left panel) two examples of fits of the
PAMELA positron fraction at high energy with a component due to DM annihilations in a 
subhalo. The fits have been obtained with sample values of the WIMP
mass and a given annihilation channel, respectively 500~GeV and 1~TeV and the 
monochromatic $e^+\, e^-$  final state. We have also assumed that the source moves with constant 
velocity $v_s = 300 \,{\rm km \,s^{-1}}$ on the trajectory perpendicular to the Galactic plane
already introduced for Fig.~\ref{fig:scal}. Looping over the propagation models
discussed in the previous Sections, we have extracted the best-fit values for the source distance
and the total annihilation rate in the source $\Gamma$. For a given propagation model, the 
background is inferred normalizing the proton flux to the locally observed spectrum, making a 
prediction for the secondary electron and positron components using the Galprop package,
and deriving the primary electron spectral index and normalization from the FERMI data. 
Including then the DM component,
the fit to the PAMELA, FERMI and HESS data (the HESS datasets are rescaled within systematic 
uncertainties to match the normalization from FERMI) is performed allowing for a 20\% variation
in the normalization of the secondary spectra (reflecting various uncertainties on their modelling)
and on the primary electrons, as well as a slight tilt in the primary electron spectral index;
the fit disregards PAMELA datapoints below 10~GeV, since the force-field method we are 
implementing to take into account solar modulation, with modulation parameter of 0.55~GV, 
is probably  not sufficiently accurate, and a modulation which takes into account the charge sign 
should be implemented instead~\cite{Adriani:2008zr}. For a given WIMP mass, and for all
propagation parameters except for the thin halo model (model B),  this procedure produces 
configurations with moderate to small $\chi^2$. They are restricted in a rather narrow range of 
allowed distances; however this range shifts with WIMP mass, choice of the propagation model 
and the velocity of the source along the orbit. 
The parameters for the two sample models obtained from the fit and plotted in Fig.~\ref{fig:pamela1} 
are given in the first two rows of Table~\ref{tab:fit}.

{\small
\begin{table}[b]
\begin{center}
\begin{tabular}{|c|c|c|c|c|c|c|c|c|}
\hline
$\quad\quad$ & $M_\chi$ & annihilation & $\Gamma$                     & ${\mc V}_s$ & prop.  & d 
& $\Phi_\gamma(E>0.1\,{\rm GeV})$                  & $\chi^2$  \\
$\quad\quad$ & GeV       & channel & $10^{36}\,{\rm s^{-1}}$ & kpc$^3$        &    model                & kpc 
& ${\rm cm}^{-2}\, {\rm s}^{-1}$ & (d.f.=50) 
\tabularnewline
\hline
\hline
1&1000 & $e^+/e^-$ & 20.9 &  $\,5.3  \cdot  10^5$ & AH & 4.25 & $1.2 \cdot 10^{-8}$ & 46.7
\tabularnewline
\hline
\hline
2& 500 & $e^+/e^-$ & 73.4 &  $\,4.6  \cdot  10^5$ & CH & 6.25 & $1.6 \cdot  10^{-8}$ & 42.3 
\tabularnewline
\hline
\hline
3& 5000 & $\tau^+/\tau^-$ & 1.9 &  $\,12.6  \cdot  10^5$ & AH & 1.54 & $1.1 \cdot  10^{-8}$ & 44.4 
\tabularnewline
\hline
\hline
4& 3000 & $\tau^+/\tau^-$ & 2.4 &  $\,5.5  \cdot  10^5$ & CH & 1.43 & $1.4 \cdot  10^{-8}$ & 60.9
\tabularnewline
\hline
\end{tabular}
\end{center}
\caption{Sample DM models }
\label{tab:fit}
\end{table}

}

\begin{figure}[t]
 \begin{minipage}[htb]{8cm}
   \centering
   \includegraphics[width=7.8cm]{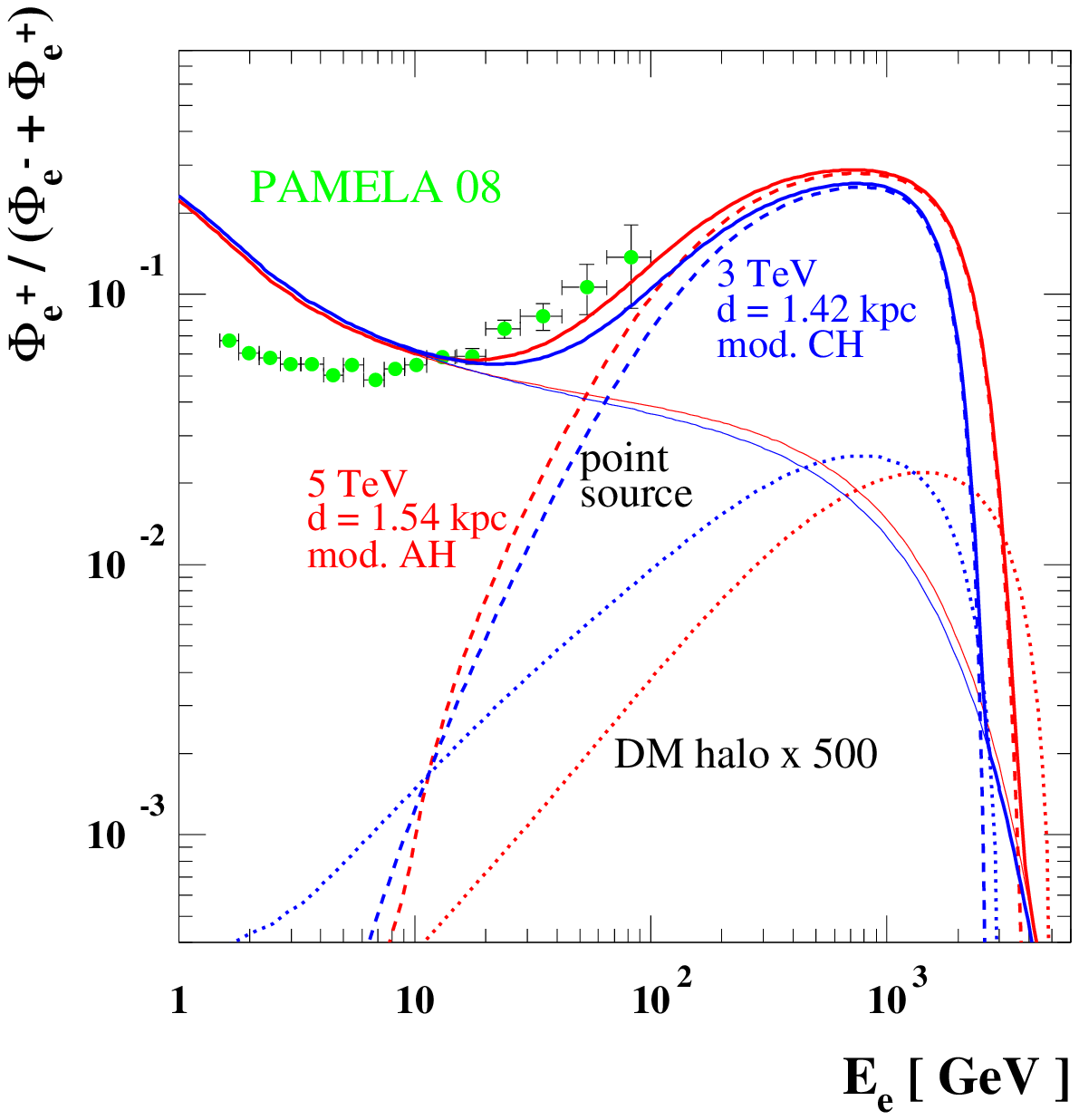}
 \end{minipage}
 \ \hspace{3mm} \
 \begin{minipage}[htb]{8cm}
   \centering
   \includegraphics[width=7.8cm]{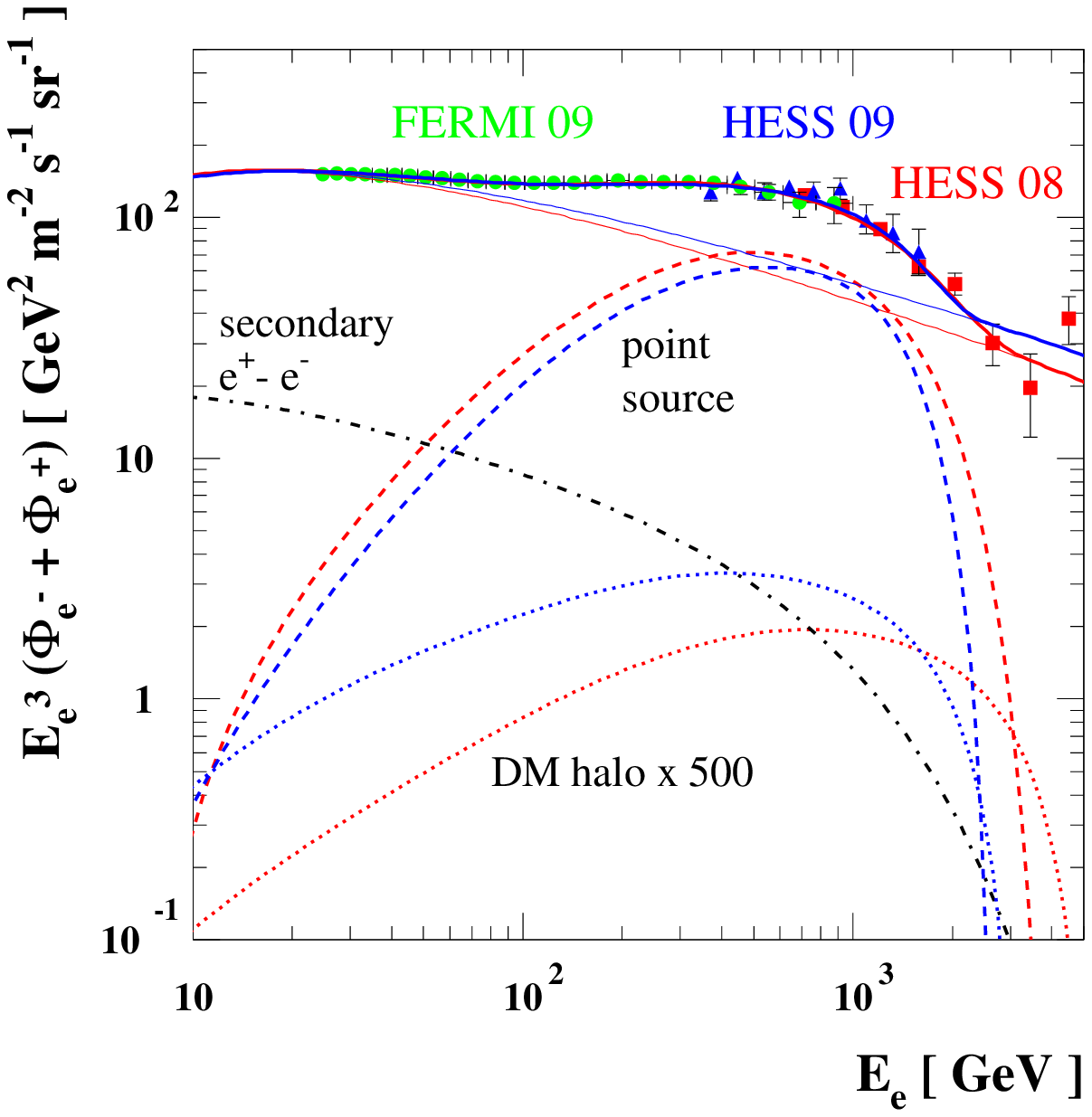}
 \end{minipage}
      \caption{Two further examples of fits of the PAMELA positron fraction (left panel) and of the sum of
     the electron and positron fluxes (right panel) with a component due to DM annihilations in
     subhalo. With respect to Fig.~\protect{\ref{fig:pamela1}}, we are now considering a different 
     annihilation channel, i.e. $\tau^+\, \tau^-$, and that the primary electron spectral index 
     is slightly harder than the one you would infer from the FERMI data.} 
\label{fig:pamela1b}
\end{figure}

Analogously, Fig.~\ref{fig:pamela1b} shows two sample fits of the data in the case of a primary 
electron spectral index slightly softer, according to the picture mentioned at the beginning of this Section. 
Since we need to account for a positron/electron exotic contribution up to 1-2~TeV, DM WIMPs needs 
to be rather heavy (e.g., we take 5~TeV and 3~TeV), while the choice of the annihilation channel 
is again not critical (we chose to refer to the case of pair annihilation into the $\tau^+\, \tau^-$ 
final state). The results from the fit are reported in the last two rows of Table~\ref{tab:fit}. 
A few issues should be stressed looking at this table. First, for all the four selected models, 
the $\chi^2$ indicate fairly good fits, although not exceptionally good (one should also take into 
account that, especially for FERMI, errors in the data-set are correlated, while we have simply 
added in quadrature statistical and systematic errors); our interest is, however, to show the feasibility 
of the framework, rather than playing with all possible uncertainties in the background and the signal
to improve the fits further (and indeed a slight readjustment of the background could 
lead to better fits in all cases). A second issue is that the values which we find for the total 
annihilation rate $\Gamma$ are very large for all models, as one sees converting them to 
the annihilation volume ${\mc V}_s$ under the hypothesis that the annihilation cross section 
is of the order  $\sigma v \simeq 3 \cdot 10^{-26}$~cm$^3$~s$^{-1}$ (this is, roughly speaking, 
the level needed for a thermal relic WIMP to match the DM density in the Universe in case 
of standard assumptions for the Universe thermal history, and without invoking an enhancement 
in the cross section going from the freeze-out to the zero temperature limit, see, 
e.g.,~\cite{Bergstrom:2000pn}). Annihilation volumes of the order of $10^5-10^6$~kpc$^3$
are much larger than typical values predicted for DM substructures in N-body simulations of 
hierarchical clustering for Milky-Way size DM halos, with a realization probability for a configuration containing these sources, supposing one can extrapolate from the results shown in Ref.~\cite{Brun:2009aj} for static sources, below few in $10^{-4}$. On the other hand, in a different scenarios, such 
as the one in which the adiabatic formation of an intermediate mass black hole
drives a sharp enhancement of the dark matter density inside the hosting 
substructure~\cite{Bertone:2005xz}, the probability density of ${\mc V}_s$ for these sources has 
a peak at about  $10^6$~kpc$^3$ and a tail extending to much larger values~\cite{Brun:2007tn}. 
There is also the possibility that the reference value  $\sigma v \simeq 3 \cdot 10^{-26}$~cm$^3$~s$^{-1}$ is a significant underestimate of the pair annihilation cross section, and in this case  
${\mc V}_s$ would be appropriately downscaled.

Having shown that it is not manifestly implausible to interpret current positron/electron data within 
the single DM substructure scenario, the examples chosen for Figs.~\ref{fig:pamela1} and \ref{fig:pamela1b} question the standard methods applied to extract model-independent informations
on the DM candidate: It is usually assumed that interpreting an excess as a DM signal gives an 
estimate of the DM particle mass when the energy threshold for the excess is detected, that
the spectral shape of the excess determines the dominant annihilation channel, while the 
normalization is mainly an indicator of the level of the pair annihilation rate. Here, instead, the 
threshold is just setting a lower limit to the DM particle mass, since the energy at which the exotic
component dies out depends also on the distance of the DM point-source. Regarding  the shape
of the spectrum, we have shown that this is a transient, keeping little memory of the spectrum
at the source, and indeed, in the plot,  the spectral shapes of the single source contribution and 
that from the smooth DM halo component are sensibly different in all examples. Finally, the level
of the induced flux depends mainly on the product $\sigma v \cdot {\mc V}_s$, and it is difficult 
to give any solid estimate for the annihilation volume ${\mc V}_s$.

\begin{figure}[t]
 \begin{minipage}[htb]{8cm}
  \centering
   \includegraphics[width=7.8cm]{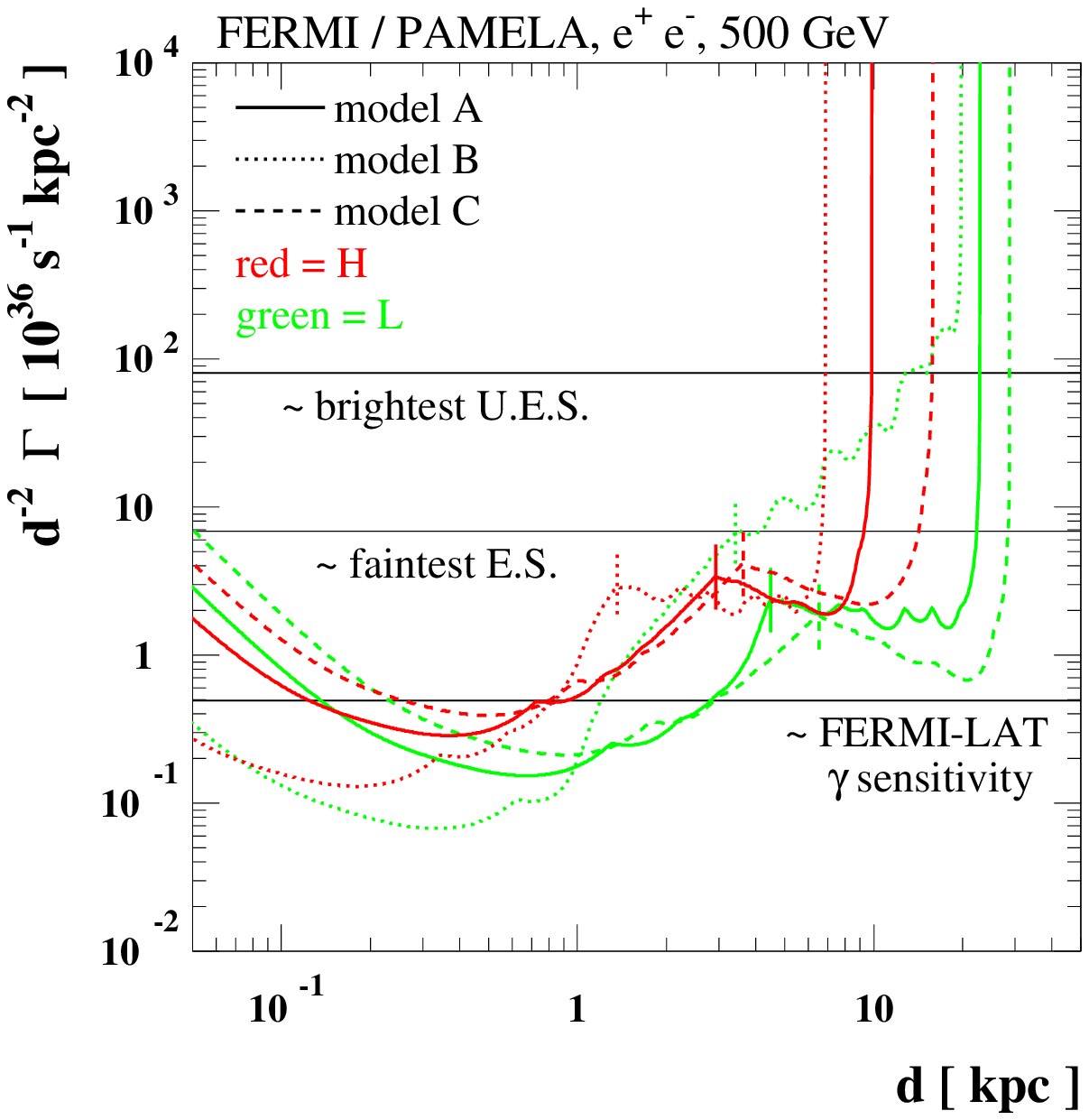}
 \end{minipage}
 \ \hspace{3mm} \
 \begin{minipage}[htb]{8cm}
   \centering
   \includegraphics[width=7.8cm]{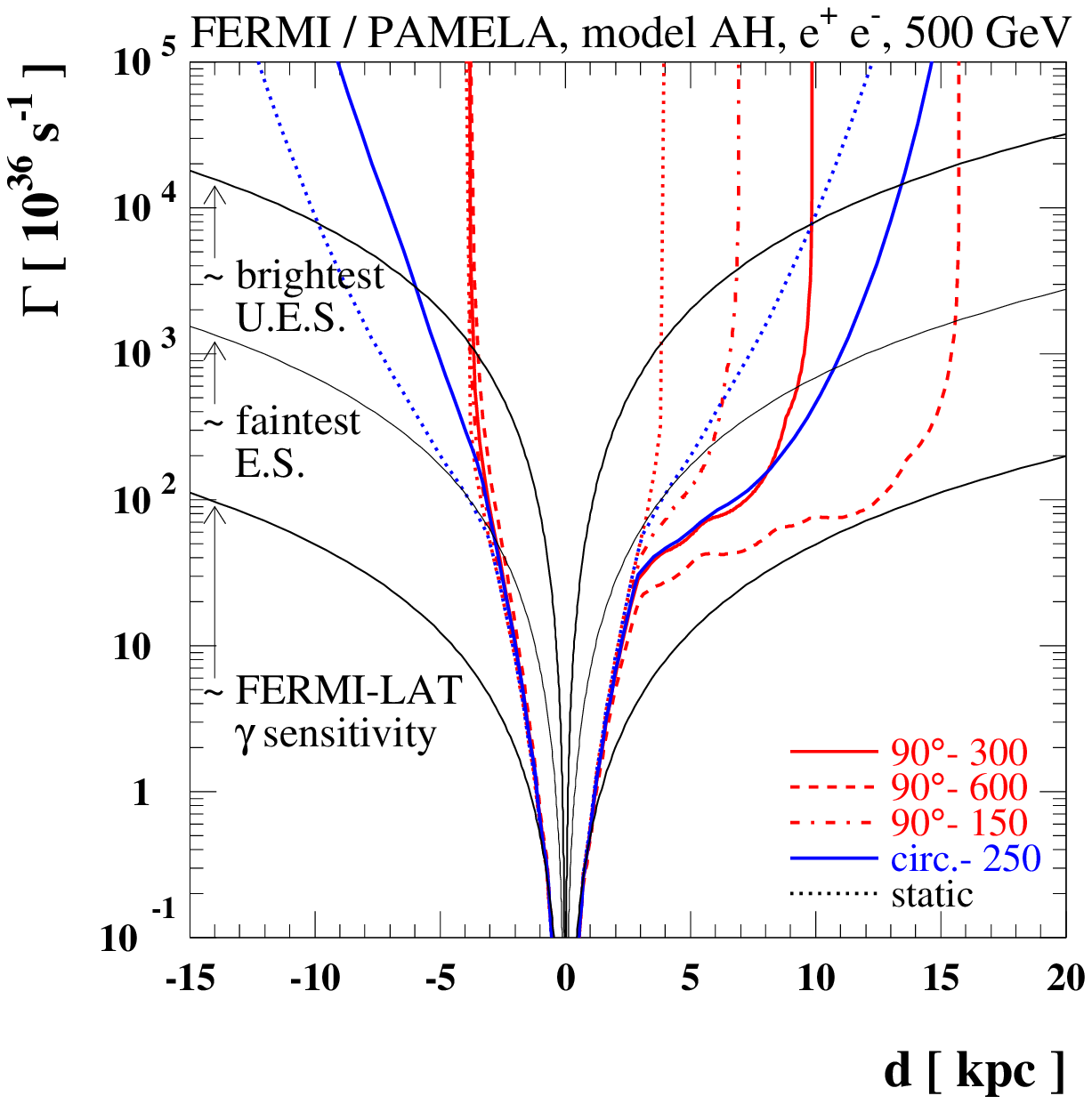}
 \end{minipage}
     \caption{Upper limits on the total annihilation rate $\Gamma$ derived from the
     PAMELA data on the positron ratio and FERMI data on the all electron flux. A sample WIMP model
     is considered ($M_\chi=500$~GeV and monochromatic $e^+\,e^-$ final state), while we loop
     over propagation model configurations (left panel) and a few possibilities for the point source 
     orbit (right panel). Results are compared to the level of $\Gamma$ required to match either 
     the brightest unidentified EGRET source, the faintest source detected by EGRET and the level
     of the $\gamma$-ray sensitivity for FERMI.}
\label{fig:epluslimit1}
\end{figure}

In Table~\ref{tab:fit}, we quote the prediction for the gamma-ray flux, 
integrated above 100~MeV, induced by the DM point-source in the four selected models: 
in the first two cases, with only
monochromatic $e^+/e^-$ as tree-level final state of annihilation, the flux is due to photons 
emitted as a final state radiation (FSR), with the rate and photon energy distribution for the 
$\chi \chi \rightarrow e^+ e^- \gamma$ which are estimated in terms of the lowest order process
$\chi \chi \rightarrow e^+ e^-$ and in the approximation for $m_e\ll M_\chi$ (see, 
e.g.,~\cite{Bergstrom:2008ag} and references therein; an eventual model-dependent  
"internal bremsstrahlung" contribution~\cite{Bergstrom:2008gr} is not included here). For the
last two cases, the annihilation into $\tau^+/\tau^-$ gives rise to decay chains containing 
neutral pions, which in turn decay to two photons; this process is accounted for by  
linking to the Pythia Montecarlo simulations as provided by the \ds\ package~\cite{Gondolo:2004sc}.
The level of the integrated gamma-ray fluxes, slightly above $1\cdot 10^{-8}$~cm$^{-2}$~s$^{-1}$,  
is much lower than the one for the brightest unidentified EGRET source, i.e. about 
$7\cdot 10^{-7}$~cm$^{-2}$~s$^{-1}$, or the level of the faintest source detected by EGRET,
i.e. about $6\cdot 10^{-8}$~cm$^{-2}$~s$^{-1}$~\cite{Pavlidou:2007su} (in a more detailed
comparison one should consider also the fact that the energy spectra for the sources proposed
here are sensibly softer than typical spectra for EGRET sources), but well within the sensitivity
for the FERMI~LAT instrument, about $4\cdot 10^{-9}$~cm$^{-2}$~s$^{-1}$ considering intermediate 
to high latitudes and a 2~year data taking period~\cite{Baltz:2008wd} (again we are not taking into
account spectral features, and also the fact that the sensitivity we quote for a DM signal was 
computed based on a gamma-ray background as extrapolated from EGRET GeV measurements of
the diffuse gamma-ray emission which are not being confirmed by FERMI and will have to be 
sensibly lowered, see, e.g.,~\cite{FERMI:preliminary}). This is a trend we see in general.

\begin{figure}[t]
 \begin{minipage}[htb]{8cm}
   \centering
   \includegraphics[width=7.8cm]{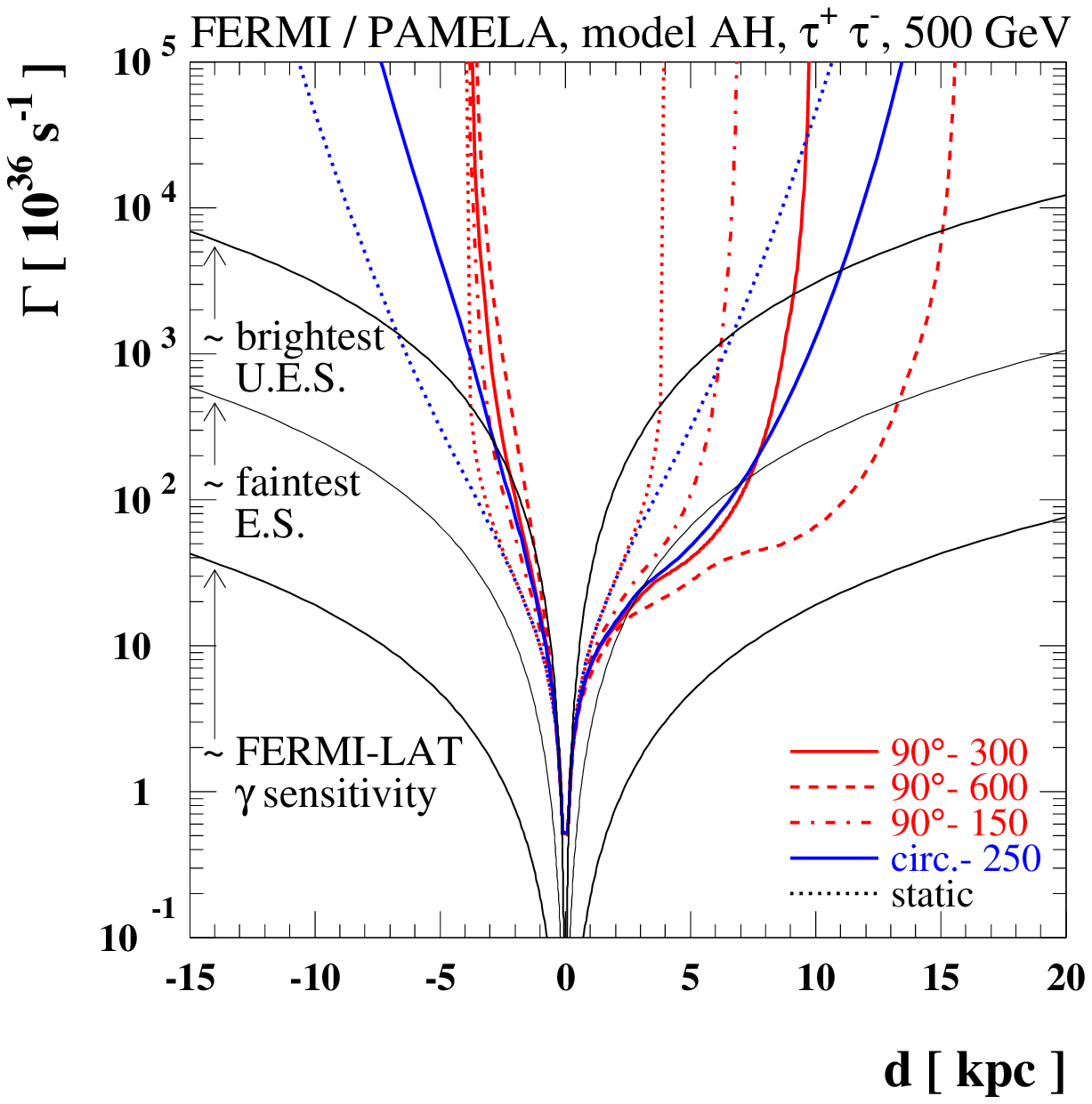}
 \end{minipage}
 \ \hspace{3mm} \
 \begin{minipage}[htb]{8cm}
   \centering
   \includegraphics[width=7.8cm]{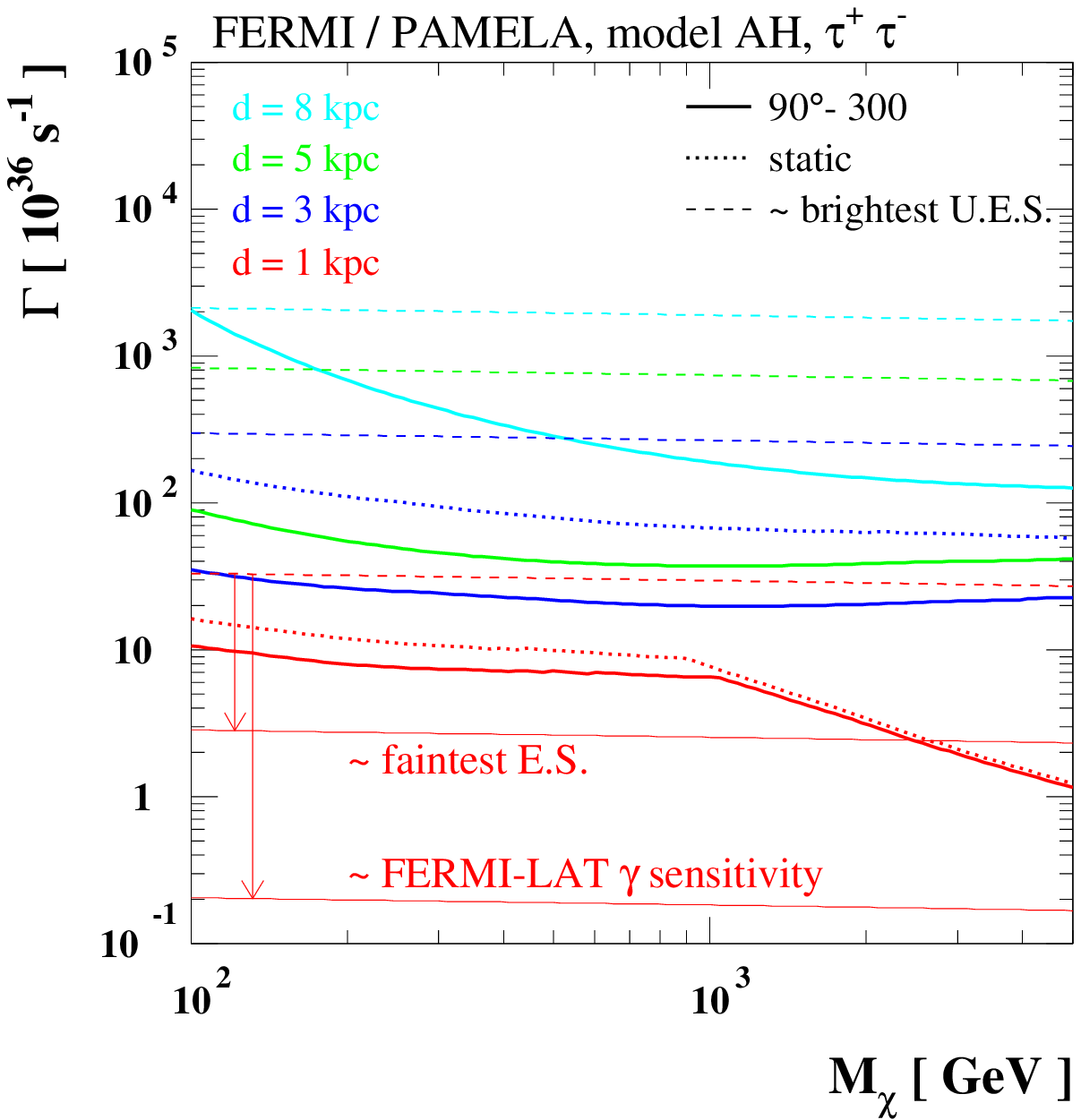}
 \end{minipage}
     \caption{Upper limits on the total annihilation rate $\Gamma$ derived from the
     PAMELA data on the positron ratio and FERMI data on the all electron flux. The left panel is 
     specular to the right panel of Fig.~\protect{\ref{fig:epluslimit2}}, but for the $\tau^+\,\tau^-$ final 
     state. In the right panel the mass scale for the DM candidate is varied and a few sample values 
     of the distance for a source moving away from the observed along a vertical orbit are considered.         
     Results are compared to the level of the $\Gamma$ required to match either 
     the brightest unidentified EGRET source, the faintest source detected by EGRET and the level
     of the $\gamma$-ray sensitivity for FERMI.}
\label{fig:epluslimit2}
\end{figure}

In Fig.~\ref{fig:epluslimit1}, 
we compute the upper limit to the total annihilation rate $\Gamma$ from the
PAMELA data on the positron ratio and from the FERMI data on the "all-electron" flux
(limits are extracted separately for the two datasets, allowing for extra freedom
in the background normalization and spectral index; the most stringent one is 
displayed). These constraints are compared to the values 
of $\Gamma$ corresponding to an integrated
$\gamma$-ray luminosity at the level of the brightest unidentified EGRET source, of the faintest source
detected by EGRET and of the FERMI $\gamma$-ray sensitivity. We have chosen a benchmark WIMP model, with
$M_\chi=500$~GeV and annihilating into monochromatic $e^+\,e^-$. We display the results
for: {\sl i)} the reference vertical orbit and different propagation models (left panel, the 
displayed cases are the same as in Fig.~\ref{fig:scal}, see the relative discussion in the text; the
values of $d$ refer to the distance of the point source from the observer at the time when the
positron flux is measured, after the source has passed in the neighborhood of the observer);
{\sl ii)} the reference propagation model AH and a few possibilities for the point-source orbit (right
panel, the sample orbits are a subset of those considered in Fig.~\ref{fig:scal2}; negative $d$ refers
to an approaching source, positive $d$ to a source moving away from the observer). In the right panel 
of Fig.~\ref{fig:epluslimit1}, we also sketch the error that can be induced by
estimating positron constraints assuming the stationary limit.
Dotted curves refer to static point-sources and show (making the comparisons with the appropriate color coding) that the bounds, except for nearby sources, are systematically overestimated for approaching sources and greatly underestimated for 
sources moving away from the observer.
Clearly, the gamma-ray luminosity of a point-source scales with $1/d^2$. 
The scaling with distance, or better with time, of the 
local positron flux depends on many different ingredients which are hard to extract from observations.
The picture, however, is intuitively clear: unless the DM point-source is extremely close to the observer inducing a very large gamma-ray flux (at a level which can be already excluded by EGRET), 
it can generate a substantial contribution to the electron/positron flux measured by 
PAMELA and FERMI, without
being in conflict with present gamma-ray observations. In particular, we expect, for intermediate 
distances, the balance to lean towards the electron/positron side since there is a time interval during which
electrons/positrons go through a diffusion transient while the gamma-ray intensity decreases as the inverse of 
time squared. When the diffusion transient is over and the local positron population 
becomes negligible, i.e. at large positive distances in the plot, the gamma-ray limit takes over again.
On the other hand, if a substantial contribution to the positron flux 
measured by PAMELA is actually due to a DM point source, it is unlikely that such source will not be 
detected by FERMI in $\gamma$-rays. This does not necessarily mean that it will be identified as a DM source. 
Indeed, FERMI will measure spectra only up to about 300~GeV and spectral features might not be easily 
identifiable, as well as the cross correlation with the positron flux is not 
anymore unambiguous, as we already stressed.

These conclusions do not depend critically on the WIMP model, as we show in Fig.~\ref{fig:epluslimit2}.
In the left panel, we consider the same configurations of the right panel in Fig.~\ref{fig:epluslimit1}, but now for a
WIMP model annihilating into $\tau^+\,\tau^-$. In this case, the $\gamma$-ray yield is enhanced (being the $\pi^0$-decay channel open) and the electron/positron yield is softer; still, at intermediate distances, a picture inducing a sizable local flux of positrons is not in conflict with the EGRET observations, but again within the FERMI $\gamma$-ray sensitivity. In the right panel of Fig.~\ref{fig:epluslimit2}, we show the constraints on $\Gamma$ obtained considering a few values for the source distance (still moving away from the observer on a vertical orbit with a velocity of 300~km~s$^{-1}$) and varying the WIMP mass. The interplay between electron/positron and gamma-ray observations is changing only slightly with mass, while 
it is again evident that the static approximation may be misleading.

 \section{Antiprotons from a dark matter point source}
\label{sec:antiproton}

\begin{figure}[t]
 \begin{minipage}[htb]{8cm}
   \centering
   \includegraphics[width=7.8cm]{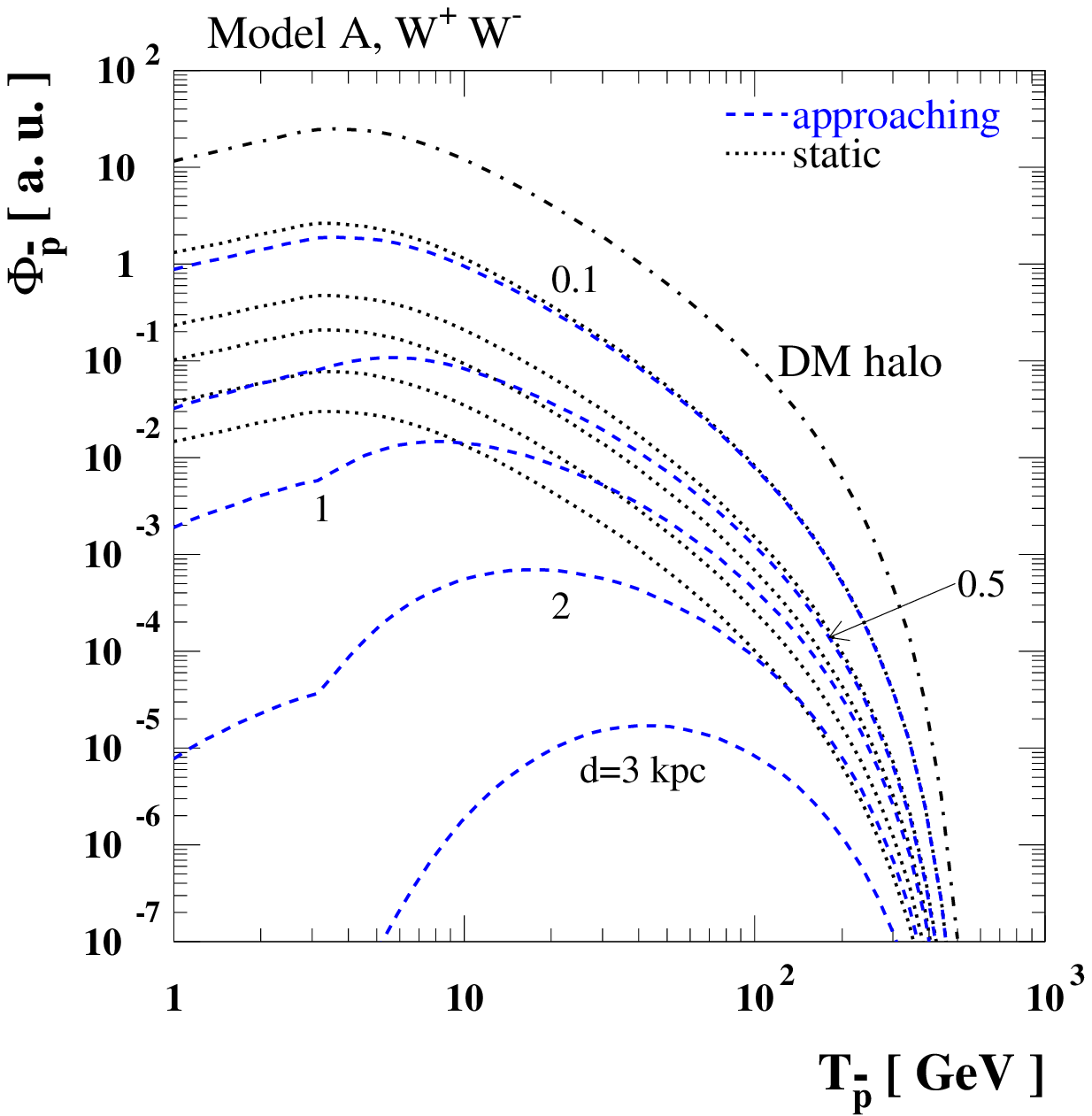}
 \end{minipage}
 \ \hspace{3mm} \
 \begin{minipage}[htb]{8cm}
   \centering
   \includegraphics[width=7.8cm]{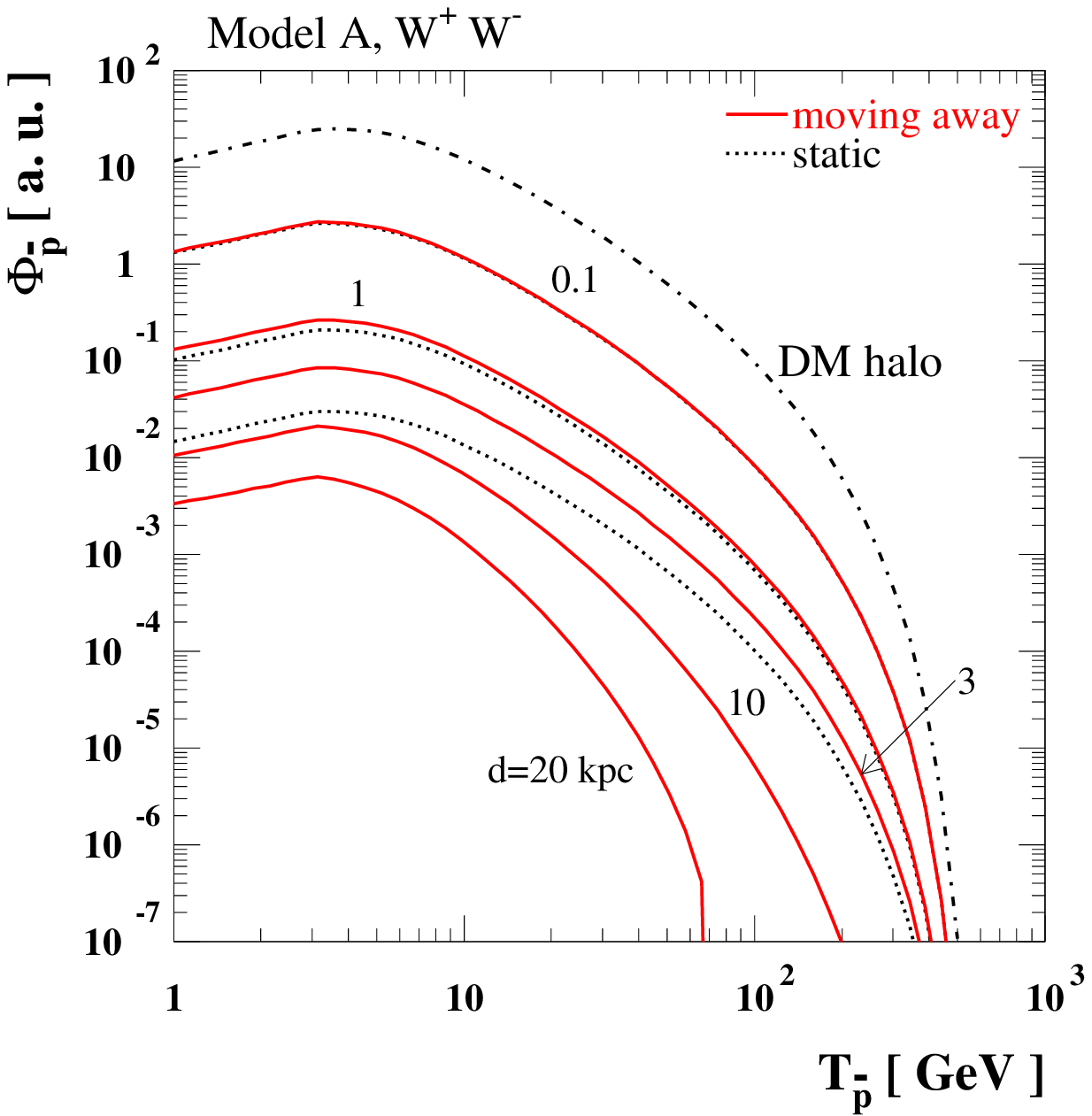}
 \end{minipage}
     \caption{Spectral shapes for the local antiproton flux due to a dark matter point source moving 
     on a vertical orbit with constant velocity $v_s = 300 \,{\rm km \,s^{-1}}$ and for different values of
     the distance $d$ of the source at the time of observation. In the left panel, the source is getting 
     closer to the observer. In the right panel, the source, after passing in the vicinity of the observer, is moving 
     away. A WIMP mass of 500~GeV and pair annihilations into  $W^+\, W^-$ have been assumed,
     as well as the propagation model~A. For comparison, spectral shapes for a static source at the 
     corresponding distances and for pair annihilations in the smooth DM halo are also shown.}  
\label{fig:pbscal1}
\end{figure}

We focus now the discussion on the possibility that dark matter pair annihilations within a substructure
give rise to an antiproton yield. This is possible if the final states of annihilation include 
weak gauge bosons or quarks, instead of leptons only, as we considered so far.
Analogously to the positron case, in the limit of a point-like dark matter substructure,
the antiproton source function takes the form:
\be
Q_{\bar{p}}(\vec{r}, t,E) = \delta^3 \left[\vec{r}-\vec{r}_p(t)\right]  \frac{dN_{\bar{p}}}{dE} \,\Gamma\,,
\label{eq:pointsourcepb}
\ee
where $dN_{\bar{p}}/dE$ is the differential antiproton yield per annihilation and $\Gamma$ is the
total dark matter annihilation rate in the source already introduced above. Solving the propagation 
equation with this source function, we find the corresponding number density per unit energy:
\be
n(\vec r,E,t) =  \frac{dN_{\bar{p}}}{dE} \,\Gamma\, \int_{-\infty}^t dt_0\; G(\vec r,t,E;\vec {r}_p,t_0)\,, 
\ee
where the Green function $G$ is given in Appendix~\ref{app:pbar}. As for positron, it is instructive 
to consider its approximate form by neglecting boundary conditions; assuming at first 
that also antiprotons annihilations in the gas disc along propagation play a minor role, $G$ goes like:
\be
G \simeq \frac{1}{\pi^{3/2} [\lambda_p(E,t,t_0)]^3}
\exp\left[-\frac{\left|  \vec r -\vec r_p(t_0)\right|^2}{[\lambda_p(E,t,t_0)]^2}\right]\,,
\label{eq:pbgreensimply}
\ee
where the diffusion length depends explicitly on time: $\lambda_p(E,t,t_0) = 2 \sqrt{(t-t_0)\,D(E)}$.
Suppose for simplicity that the substructure is moving along a straight  line trajectory with 
constant velocity $\vec{v}_s$, namely $\vec{r}_p(t_0) = \vec{r}_t + \vec{v}_s(t_0-t)$; in this 
case $G$ can be rewritten in the form:
\be
G \simeq \frac{1}{8 \pi^{3/2} \, D^{3/2} \, (t-t_0)^{3/2}}  
\exp\left[-\left(\frac{t_s}{t-t_0}-2\,\sqrt{\frac{t_s}{t_d}}c_\theta+\frac{t-t_0}{t_d}\right)\right] 
\label{eq:pbgreensimply2}
\ee
where we have introduced a "static" timescale $t_s \equiv d_t^2/4 D$, with 
$d_t=\left|  \vec r_t -\vec r\right|$ being the distance between the point-source and the observer at the time of 
observation (which would be, of course, the distance between the point-source and the observer at all times in 
the static limit), and a "dynamical" timescale $t_d \equiv 4 D/|\vec{v}_s|^2$, and we have defined
$c_\theta$ as the cosine of the angle between the vectors $\vec{v}_s$ and $ \vec r_t -\vec r$
(which is negative for a source approaching the observer and positive for those moving away). 
We recognize that $t_d$ and $t_s$ are the appropriate quantities to compare for a first 
guess on whether proper motion is relevant or not in estimating antiproton fluxes from point 
sources; in particular, Eq.~(\ref{eq:pbgreensimply2}) can be integrated over time, giving:
\be
\int_{-\infty}^t dt_0\; G \simeq \frac{1}{4 \pi \, D \, \left|  \vec r -\vec r_t\right|}  
\exp \left[ -\frac{2 t_s}{t_d} + 2 \,\sqrt{\frac{t_s}{t_d}}c_\theta \right]\,.
\label{eq:pbgreensimply3}
\ee
In the limit for $t_d \gg t_s$, this expression correctly reduces to the Green function of the
three dimensional Laplacian, while for $t_d \sim 2\,t_s$ we see that proper motion starts 
to become important. For typical values of the propagation parameters in the model, we find:
\be
\frac{t_d}{2\,t_s} = \frac{8 D^2(E)}{d_t^2 v_s^2} \simeq 1.5
 \frac{D_1^2\,E_{10}^{2 \delta} \, 10^{2 \delta-1.2}}{d^2_{t,1} v_{300}^2}\,,
\ee
where $E_{10}$ is antiproton energy in units of 10~GeV and $d_{t,1}$ is the source distance in units
of 1~kpc. As for the positrons, we find the effect of proper motion to be negligible 
either in the limit of nearby sources or going to high energies, while we expect it 
to be significant in the other cases. 

Actually, Eq.~(\ref{eq:pbgreensimply}) and (\ref{eq:pbgreensimply2}) contain an oversimplification,
since they were derived neglecting antiprotons annihilations during propagation, an effect which
defines a third timescale that could also be relevant: depending on energy, the antiproton loss via 
annihilation can be much larger than the leakage from the boundaries of the diffusion region.
When including this extra timescale, the expression for $G$ becomes less transparent; we show 
instead results for a few sample cases obtained by implementing the exact Green function.
In Fig.~\ref{fig:pbscal1}, we plot the spectral shape for the local antiproton flux from a source moving 
along the reference trajectory, namely a
path perpendicular to the Galactic plane and intersecting it at a short distance from the 
observer, with the source moving at constant velocity $v_s = 300 \,{\rm km \,s^{-1}}$.
We have also chosen the propagation model~A, and defined the dark matter model through
a sample value for the WIMP mass,  $M_\chi=500$~GeV, and assuming that the annihilation
is dominantly into $W^+\, W^-$ pairs. As in the previous plots, $d$ refers to the distance of the 
source from the observer at the time of observation and is used instead of a time variable
to compare more easily to the static limit (dotted curves in the plot). 
There is a evident transient at small and intermediate energies, soon after the source has 
entered the diffusion region; the spectrum starts to fully match the static limit case only when the 
source arrives very close to the observer. Moving away from the observer, the scaling with 
distance is less severe than in the static limit and a contribution to the local antiproton flux
persists much later than the time at which the source has left the propagation region (the vertical 
boundary is at $h_h = 4$~kpc in model~A). Since high energy antiprotons diffuse more efficiently,
at late times the spectral shape becomes steeper and starts again to differ sensibly from the shape 
of the component due to pair annihilations in the smooth dark matter halo (dash-dotted line in 
the plot). Note that the latter, contrary to the positron case, traces rather closely the shape of static 
point-sources.

\begin{figure}[t]
 \begin{minipage}[htb]{8cm}
   \centering
   \includegraphics[width=7.8cm]{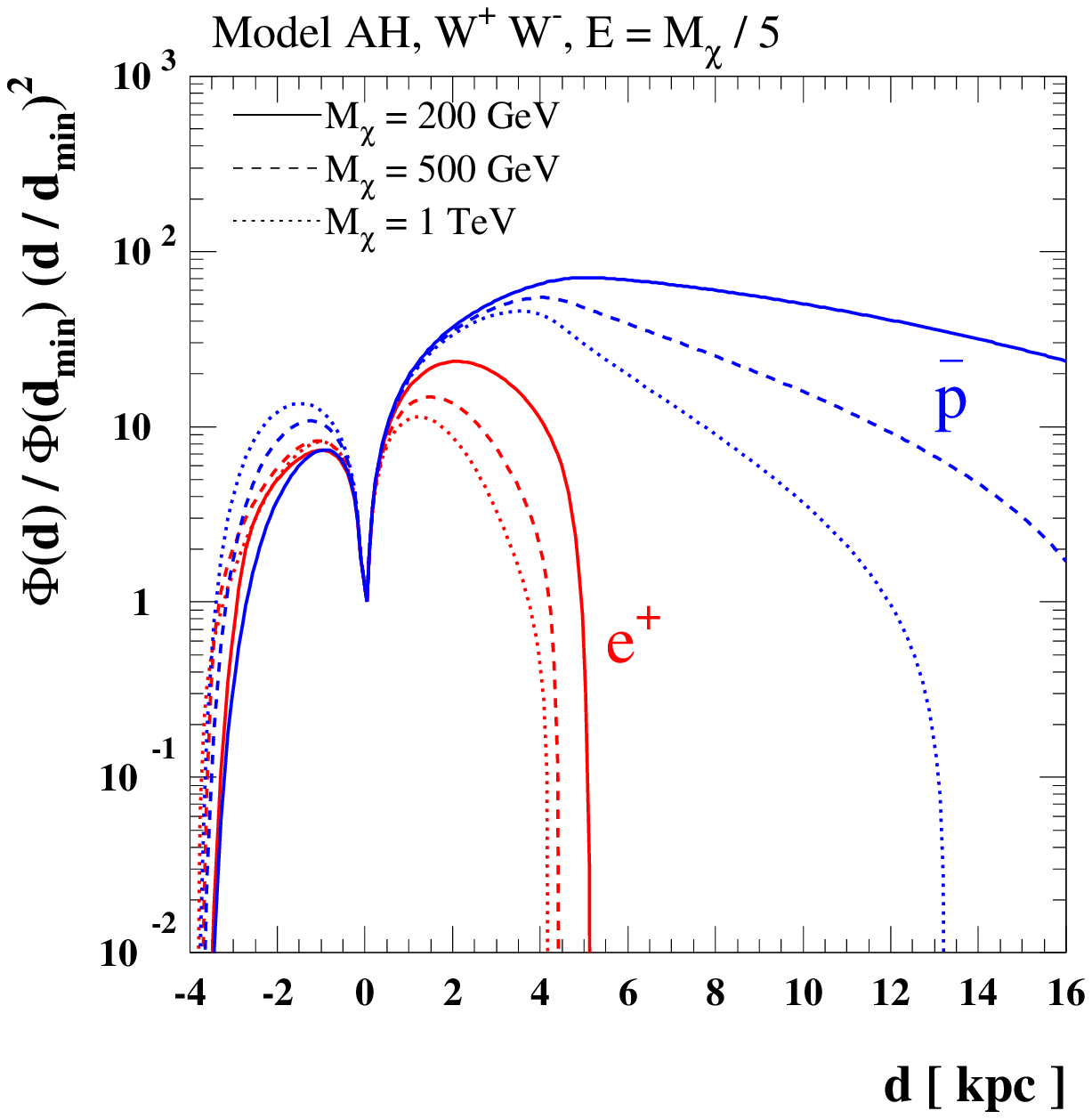}
 \end{minipage}
 \ \hspace{3mm} \
 \begin{minipage}[htb]{8cm}
   \centering
   \includegraphics[width=7.8cm]{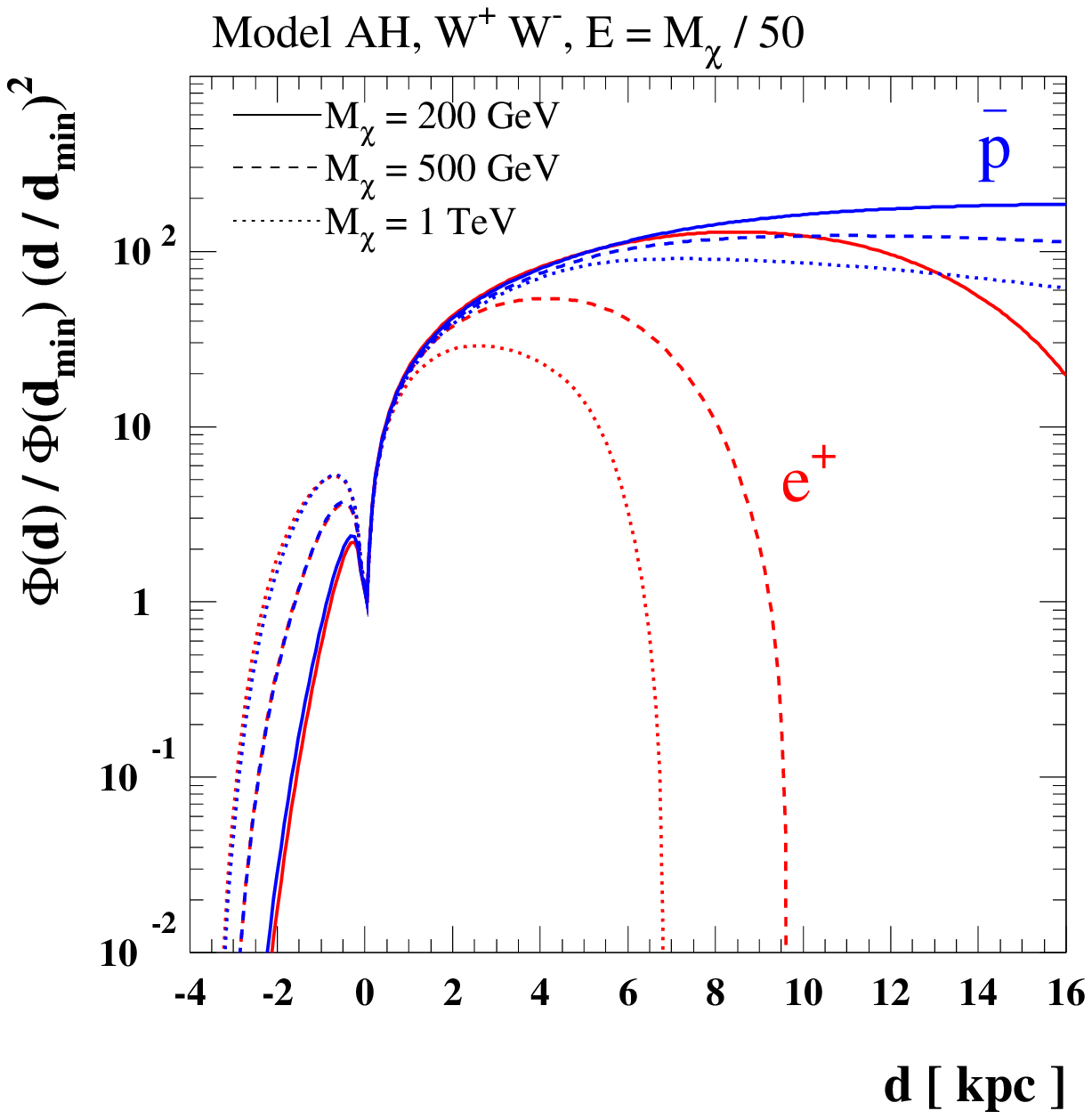}
 \end{minipage}
     \caption{The locally measured antiproton and positron fluxes as  a function of the distance 
     of the point source (negative distances label an approaching source, positive values a source 
     moving away) normalized to the fluxes at a closest encounter distance $d_{min} = 50$~pc 
     and to the scaling with distance of the companion induced gamma-ray flux, i.e. $1/d^2$. 
     $W^+\, W^-$ annihilation channel is considered (giving rise at the same time to antiprotons,
     positrons and photons) and three sample values of the WIMP mass $M_\chi$. The antiproton
     and positron fluxes are plotted at the energies $E$ which are equal to $M_\chi/5$ (left panel)
     and $M_\chi/50$ (right panel).}
\label{fig:pbscal2}
\end{figure} 

\begin{figure}[t]
 \begin{minipage}[htb]{8cm}
   \centering
   \includegraphics[width=7.8cm]{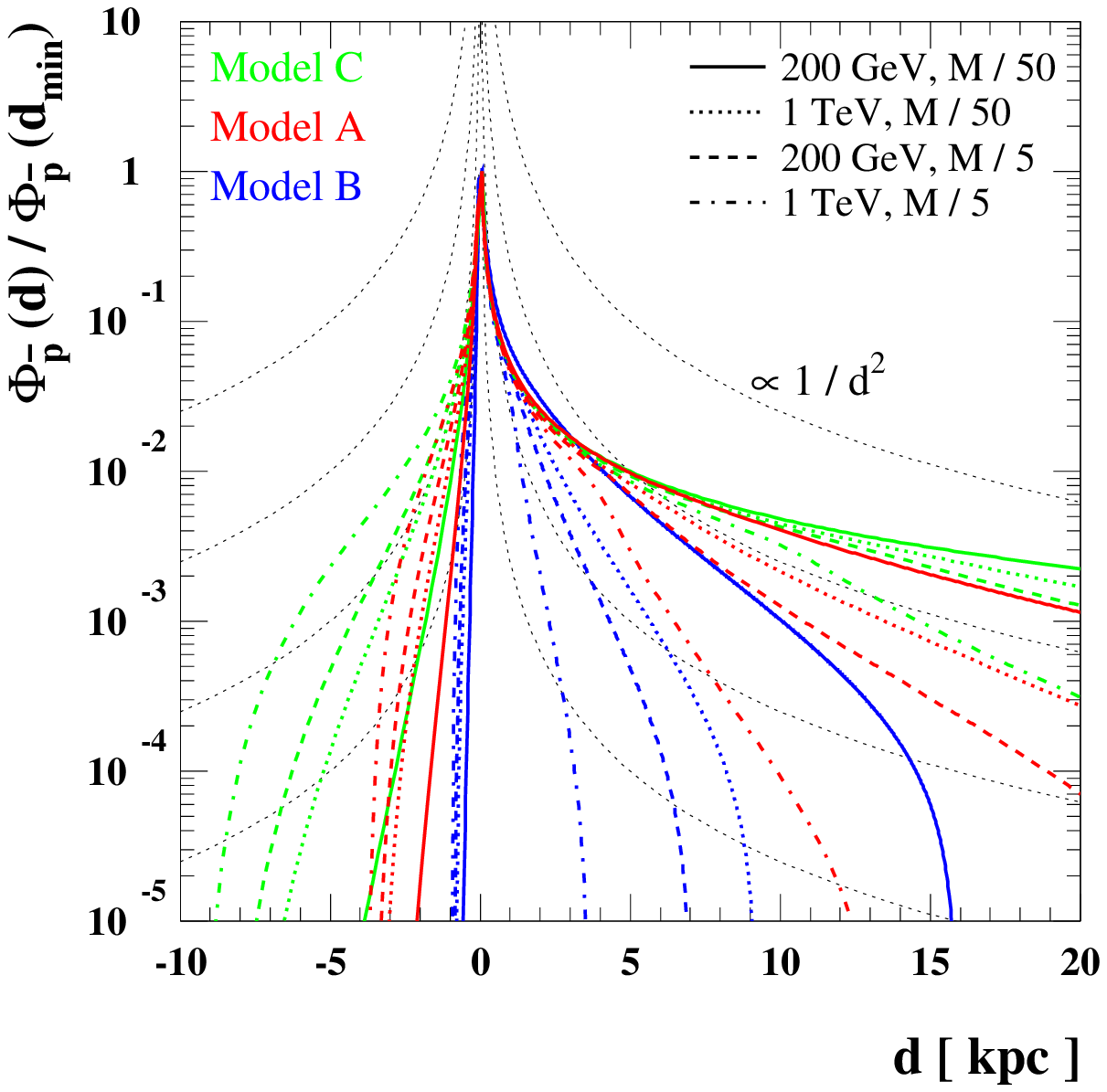}
 \end{minipage}
 \ \hspace{3mm} \
 \begin{minipage}[htb]{8cm}
   \centering
   \includegraphics[width=7.8cm]{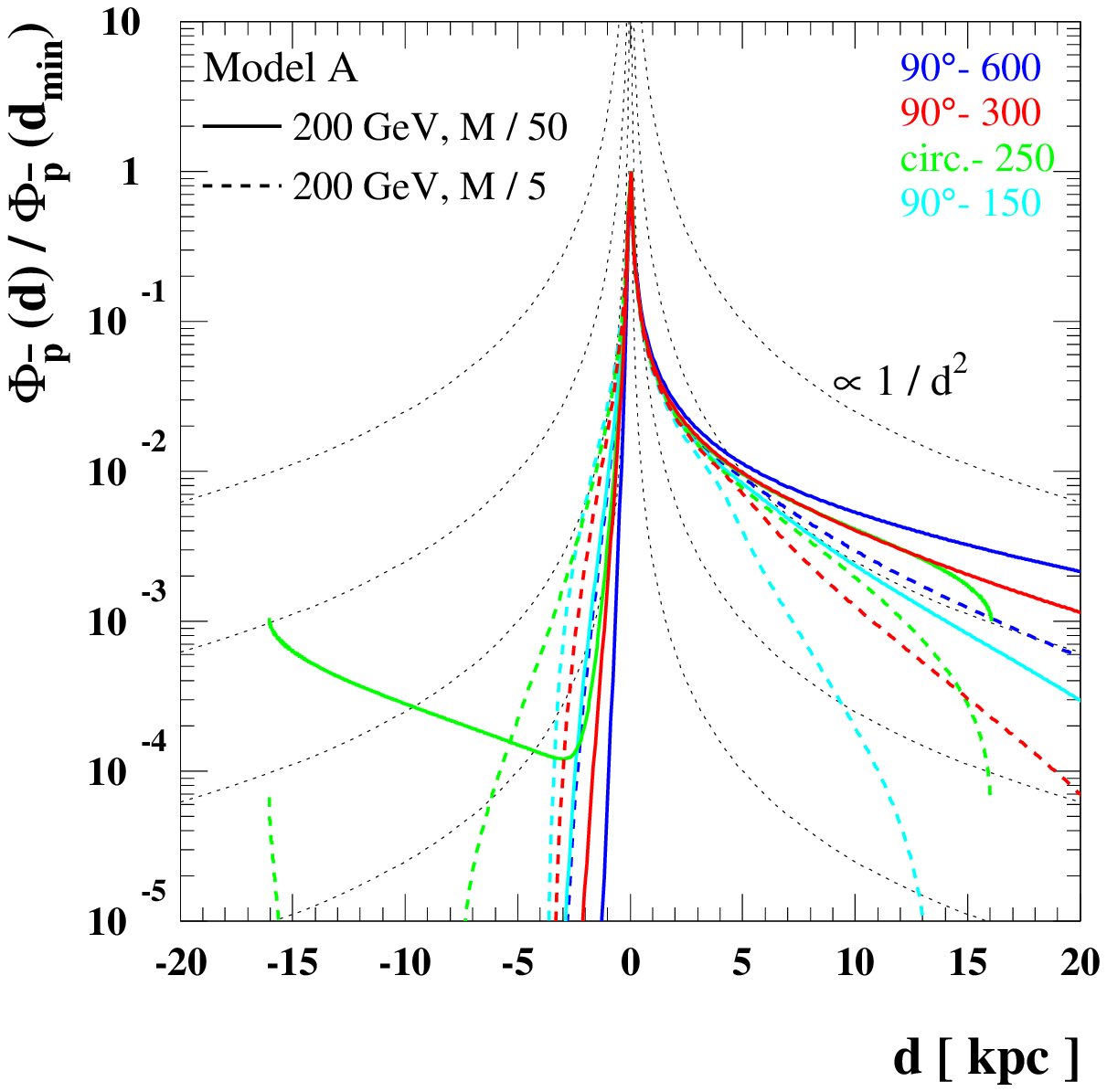}
 \end{minipage}
     \caption{Scaling of the locally measured antiproton flux with the distance of the point source
     from the observer (negative distances label an approaching source, positive values a source 
     moving away), for two WIMP masses  $M_\chi$ (200~GeV and 1~TeV) and two sample energies 
     $E$ (equal to $M_\chi/5$ and $M_\chi/50$), and assuming  
     pair annihilations into $W^+\, W^-$. In the left panel, the results refer to the standard vertical 
     trajectory and to our three benchmark propagation models A, B \& C (intermediate,
     thin-halo \& thick-halo, see text for details). In the right panel, for the propagation model~A,
     the color coding refers to different orbits, namely three vertical trajectories with velocities of the
     source changing from 300 to 600 or 150~km~s$^{-1}$, and a circular orbit in the Galactic plane
     with velocity matching the local circular velocity 250~km~s$^{-1}$. } 
\label{fig:pbscal3}
\end{figure}

The $W^+\, W^-$ annihilation channel is a copious source of positrons and gamma-rays 
(mainly from the production and decay of pions). In Fig.~\ref{fig:pbscal2}, we compare
the scaling with distance (i.e. time) of the locally measured antiproton and positron fluxes at an
energy corresponding to a fraction of the WIMP mass (1/5 and 1/50, respectively, in the left and right
panels) for three sample values of  $M_\chi$.  Both antimatter fluxes are normalized to the scaling with distance
of the induced gamma-ray emission, which goes simply like $1/d^2$.  The ratio between charged-particle
fluxes to the gamma-ray flux increases rapidly around $d=0$ since the local number density of 
positrons and antiprotons decreases less severely when the distance from the source is smaller 
than the corresponding diffusion length. At negative $d$, i.e. for a source approaching the observer, one can see again the transients due to the fact that the source has just entered the diffusion region;
these transients are essentially specular in the two cases. On the other hand,
the contribution to the local antiproton population tends to survive much longer than for the positron
component. When the source is moving away, the antiproton flux scales, at least at intermediate distances and except for very energetic antiprotons, like $1/d^2$; positrons have instead much sharper transients, due to the efficient energy 
loss term. 

Fig.~\ref{fig:pbscal2} has been obtained for the reference propagation model (AH, with the
"H" labeling the energy loss configuration for positrons only)  and the reference trajectory. As for 
the positrons, we expect the results for the antiproton flux to depend on the choice of the 
propagation parameters and the orbit for the point source. This is illustrated in 
Fig.~\ref{fig:pbscal3}: in the left panel, we show the scaling of the locally measured antiproton
flux with distance considering two sample WIMP masses, two sample energies and the benchmark
trajectory, while  looping over the three reference propagation models. The main effect here is due to the 
increase in the diffusion coefficient going from model~B to model~A and then to model~C, as well as
to the boundary conditions, entering more critically for the thin-halo model~B ($h_h=1$~kpc); note, 
however, that even for model~B a sizable contribution to the local flux may be due to sources
which are well outside the diffusion region and that would not be included in the static limit treatment.
In the right panel of Fig.~\ref{fig:pbscal3}, we consider instead a few of the orbits that we have already
introduced to discuss positron fluxes, with the pattern for source velocities which is analogous here;
note that in the circular orbit case the transient following a close encounter can  be sufficiently 
long lasting for the induced antiproton population to persist up to the next close encounter (what we 
plot is the "equilibrium" configuration after a couple of full orbits).

\begin{figure}[t]
 \begin{minipage}[htb]{8cm}
   \centering
   \includegraphics[width=7.8cm]{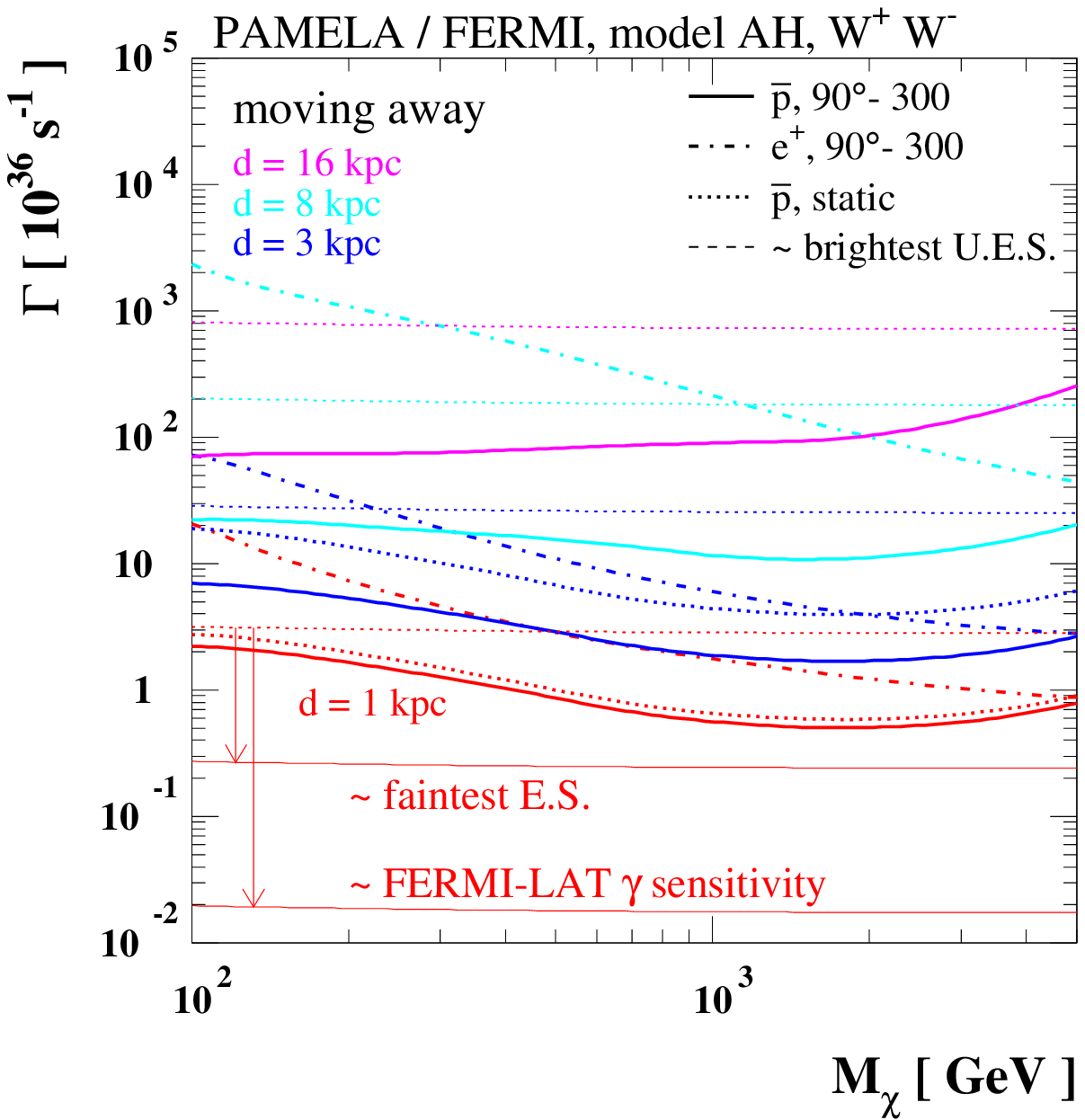}
 \end{minipage}
 \ \hspace{3mm} \
 \begin{minipage}[htb]{8cm}
   \centering
   \includegraphics[width=7.8cm]{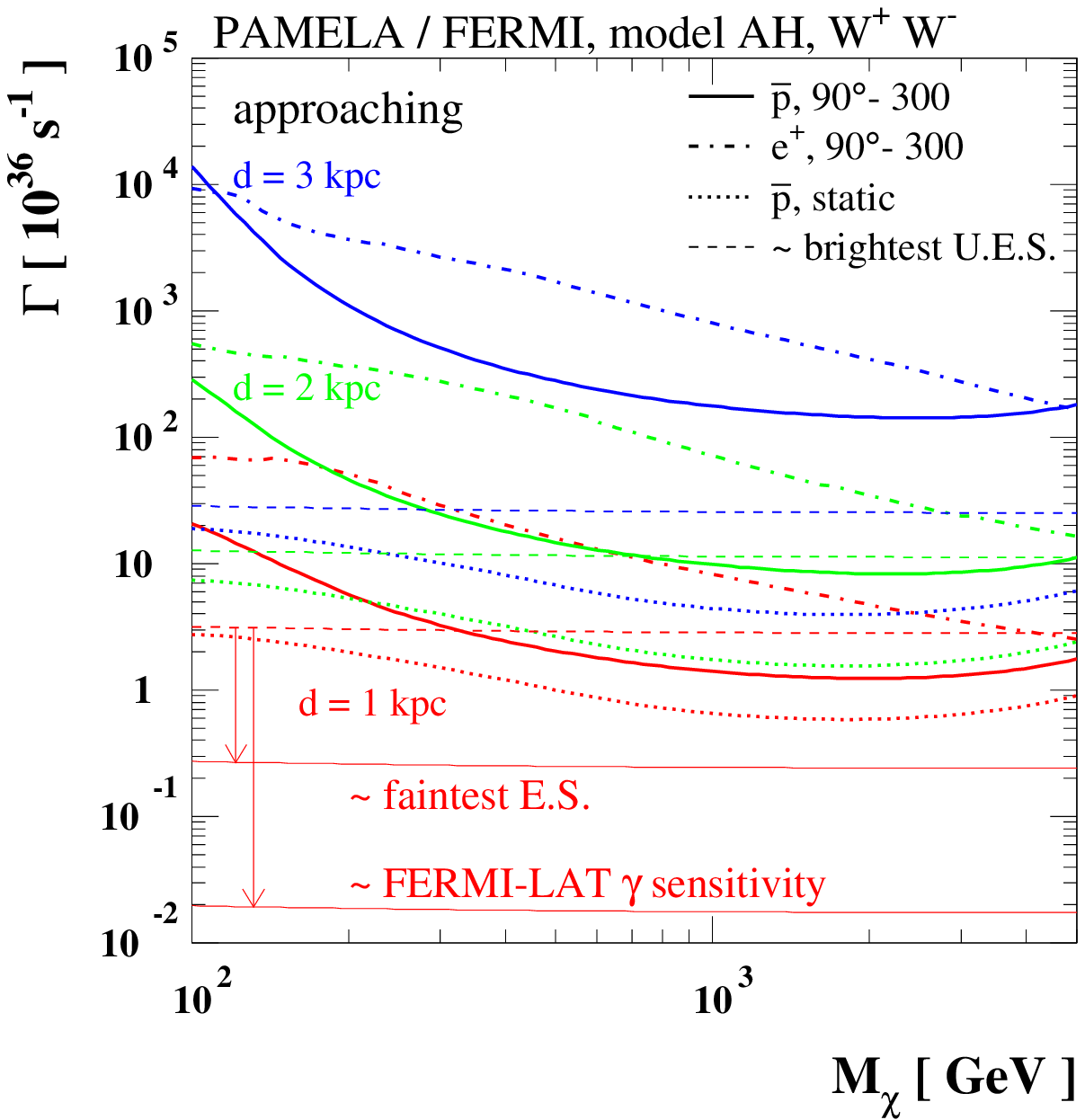}
 \end{minipage}
     \caption{Upper limits on the total annihilation rate $\Gamma$ derived from the
     PAMELA data on the antiproton fraction (solid lines), PAMELA data on the
     positron ratio and FERMI data on the all electron flux (dash-dotted curves). The source is 
     moving along the reference vertical trajectory and is approaching the observer (left panel)
     or moving away from the observer (right panel). 
     Results are compared to the level of the $\Gamma$ required to match either 
     the brightest unidentified EGRET source, the faintest source detected by EGRET and the level
     of the $\gamma$-ray sensitivity for FERMI.}
\label{fig:pbscal4}
\end{figure}

Finally, in Fig.~\ref{fig:pbscal4}, we summarize the picture by comparing, for a WIMP model
with $W^+\, W^-$ annihilation channel, the constraints on the total annihilation rate which are
set by the latest measurements by PAMELA of the antiproton to  proton ratio~\cite{Adriani:2008zq},
the electron/positron measurements by PAMELA and FERMI, as well as searches for point 
$\gamma$-ray sources by EGRET. Projected limits for the FERMI $\gamma$-ray telescope are also shown. 
Analogously to the case of the smooth DM halo component, 
we find for a DM substructure contribution that if WIMP annihilates in a channel in which antiprotons 
are copiously emitted, the  measured antiproton flux sets tighter constraints than 
electron/positron data, except possibly for extremely heavy DM candidates. This trend is reinforced 
going to larger distances. The information from EGRET is inconclusive, while FERMI
is going to cover most, if not all, the parameter space currently probed by antiproton or positron
searches.  The displayed results refer to a source moving along the reference vertical trajectory,
approaching or moving away from the observer; however, analogous conclusions hold for other
configurations. The extension to other WIMP annihilation channels also leads to similar results.

\section{Extra features of positron/electron emission for a DM point source}
\label{sec:extra}

\subsection{A gamma-ray component from inverse Compton emission on starlight}

Radiative emissions are unavoidably associated to electron/positron yields; in particular,
the Inverse Compton (IC) radiation component due to the interaction of high energy 
electrons on optical starlight is peaked in the range between tens to hundreds of GeV and may be
sizable for the intense dark-matter lepton emitters we have been considering.  

The IC photon emissivity is obtained by folding, at any given point within the diffusion region, 
the IC emission power with the electron/positron number density, see, e.g.,~ \cite{Rybicki}:
\be
j_{IC}(\nu,\vec{r},t)=2\int^{M_{\chi}}_{m_e}dE\, P_{IC}(\vec{r},E,\nu)\, n(\vec{r},E,t)\;,
\label{eqjIC}
\ee
where the IC power is given by:
\be
P_{IC}(\vec{r},E,\nu) = c\,h\nu \int d\epsilon\, \frac{dn_\gamma}{d\epsilon}(\epsilon,\vec{r})
\,\sigma(\epsilon,\nu,E) v 
\label{eqPIC}
\ee
where $\epsilon$ is the energy of the target photons, $dn_\gamma/d\epsilon$ is their differential energy
spectrum and  $\sigma$ is the Klein--Nishina cross section. The spatial dependence and spectrum of the number density of starlight photons can be computed from photometric maps of the Galaxy; we will adopt the model implemented in the public release of the Galprop numerical package~\cite{Strong:1998pw}. On the other hand, the analytic solution introduced in Section~\ref{sec:positronpoint} for computing the electron/positron number density $n$ was obtained assuming a spatially constant electron/positron energy loss term, and, hence, also a mean value for the background starlight density, rather than its value as a function of the spatial coordinates. This should not be regarded as a severe inconsistency, since the IC on starlight is just one of the effects contributing to the energy loss term, which main uncertainty is related to synchrotron emission on background magnetic fields. Having computed the IC emissivity field as in Eq.~\ref{eqjIC}, the relative $\gamma$-ray flux for a local observer is obtained simply by summing contributions along the line of sight.

\begin{figure}[t]
   \centering
   \includegraphics[width=9.cm]{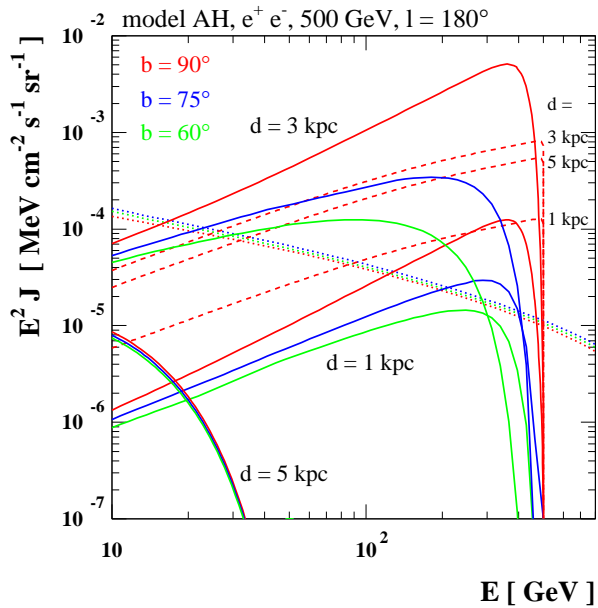}
   \caption{$\gamma$-ray spectra for a DM point-source composed by 500~GeV WIMPs annihilating into    
   monochromatic electrons and positrons, and moving along the reference vertical trajectory. 
   The solid lines refer to the IC emissivity and are computed for three different directions of observation
   labeled by the latitude $b$ (the longitude is $l=180^\circ$); 
   dashed lines refer to the FSR component in the direction of the source, 
   while dotted lines show the Galactic $\gamma$-ray background. The DM signals are computed for three 
   sample distances of the source, assuming that, for each distance, the normalization of the 
   total annihilation rate in the source is the maximum allowed by current PAMELA and FERMI 
   electron/positron data, see the limits derived in Fig.~\protect{\ref{fig:epluslimit1}}.
   }
\label{fig:ic}
\end{figure}

An example is given in Fig.~\ref{fig:ic}, where we refer again to the model introduced at the beginning of
the discussion, namely a point source composed by 500~GeV WIMPs annihilating into monochromatic 
electrons and positrons and moving on a vertical orbit intersecting the Galactic plane at a short distance from the observer, see Figs.~\ref{fig:scal} and \ref{fig:epluslimit1}. The propagation scenario considered is the AH model (recall that in this model the vertical scale height of the diffusion region is 4~kpc) and the time variable is again replaced by the distance from the observer. Having assumed a velocity of 
the point source of 300~km~s$^{-1}$, we consider three sample cases, i.e. $d=1,\, 3,\, 5$~kpc with the source that has passed close to the observer and is now moving away. The induced IC flux (solid lines) is shown for three different directions of observation, all at the longitude $l=180^\circ$ (i.e. opposite to the Galactic center), with latitude $b=90^\circ$ (the direction towards the point source in our example), $b=75^\circ$ or $b=60^\circ$. We plot also the background level corresponding to the emission from Galactic cosmic-rays (dotted lines) computed with Galprop, and the DM-induced $\gamma$-ray component due to FSR at the source (which is obviously present only in the direction of the DM substructure). The normalization of the total annihilation rate, or of the annihilation volume, has been chosen by saturating the current limits from PAMELA or FERMI, see the results in Fig.~\ref{fig:epluslimit1}. This explains the relative strength of the FSR components. 
One can see that the IC flux in the direction of the source can be as large as the FSR contribution (for $d=1~kpc$) or even larger (for $d=3$~kpc). Most notably, the spatial size of the IC emission is significant, inducing a signal larger than the background even tens of degrees away from the source. Consider, however, that the case displayed here is the most favorable, since the source is located in the portion of the sky with the faintest Galactic background, and we have not included the uncertain extragalactic component. For $d=5$~kpc, i.e. for the source outside the diffusion region, the IC term drops dramatically. This is due to the fact that there is no fresh source of high energy electrons and positrons, and the ones injected along the transit of the source through the diffusion region have rather rapidly degraded their energy. Therefore, there is no efficient high energy source to up-scatter starlight photons to $\gamma$-ray frequency. The trend is probably exaggerated by the fact that we are assuming a sharp boundary for the propagation region, while a smoother boundary condition would be certainly more physical and would probably lead to a smoother transition between the cases, $d=3$~kpc to the $d=5$~kpc. 

From the sample case discussed, we can infer that it is rather likely that if the local positron flux receives a sizable contribution from a single point-source, the induced IC flux is a relevant target for the FERMI $\gamma$-ray telescope. A more thorough discussion about this point is delayed to an upcoming dedicated analysis.

\subsection{Dipole anisotropy in the electron/positron spectrum}  

As pointed out by recent analyses (see e.g.,~\cite{Buesching:2008hr,Hooper:2008kg,Grasso:2009ma,Cernuda:2009kk}), a single nearby point-like source (e.g., a pulsar), being the dominant local positron source at high energy, induce an anisotropy which could be at a detectable level. In this Section, we reconsider this possibility in the context of DM substructure. We take into account proper motion effects, which are not relevant for bursting-like sources as pulsars. We assume the dominant contribution to the anisotropy to be given by the dipole term.
In order to detect an anisotropy along a certain direction at a good confidence level, the required number of events has to be very large. Therefore, although the anisotropy in the positron spectrum would be higher than in the total electron plus positron spectrum, we refer to the latter, being the detection prospects for FERMI much more promising than for PAMELA.

If the anisotropy is dominated by the dipole term, the intensity at a given point as a function of the direction of observation has only one maximum. The degree of dipole anisotropy can be defined by the quantity $\delta=(I_{max}-I_{min})/(I_{max}+I_{min})$, where $I_{max}$ and $I_{min}$ are the maximum and minimum $e^++e^-$ intensity with respect to direction.
Expanding the intensity in spherical harmonics up to the dipole term, we have: $I(\theta,\phi)\simeq \bar I+\delta\, \bar I \cos{\theta}+\sin{\theta}\,(I^{-1}_1e^{-I\phi}+I^1_1 e^{I\phi})$, where $\theta$ is the angle with respect to the direction of maximal anisotropy, $\bar I=(I_{max}+I_{min})/2=(I(\theta=0)-I(\theta=\pi))/2$, and $I^{-1}_1$ and $I^1_1$ are angle-independent coefficients.
Note that contrary to the case of a stationary point-source (with homogeneous and isotropic propagation parameters), the $\phi$-dependent terms of the dipole are not null. 

Although the direction of maximal anisotropy (chosen, e.g., as the $z$-axis) is, in general, unknown, we can define the particle flux $F$ along such direction by integrating over all the possible directions the projection over $z$ of the intensity, i.e., $F\simeq \int d\Omega\, I\cos{\theta}\simeq 4/3\,\pi\, \delta\, \bar I$, and by estimating it as $F\simeq D\,|\nabla n|$. The latter is obtained in the diffusion approximation~\cite{Berezinskii:1990}, but it is valid at first order (i.e., for small anisotropies) also when including energy losses. Moreover, we assume the anisotropies in directions orthogonal to $z$ to be subdominant (i.e., $\partial_z n \simeq \nabla n$). By equating the two expressions for $F$, one gets~\cite{Ginzburg:1964}:
\be
\delta=\frac{3\,D}{c}\frac{|\nabla n_{tot}|}{n_{tot}}=\frac{3\,D}{c}\frac{|\nabla n_{DM}|}{(n_{DM}+n_{CR})} \;,
\label{Eq:anis}
\ee
where $n_{DM}$ denotes the contribution to the $e^++e^-$ number density induced by a DM substructure, $n_{CR}$ is the contribution from cosmic-rays, and $n_{tot}=n_{DM}+n_{CR}$.

Neglecting boundary conditions, $|\nabla n_{DM}|$ is given by the gradient of Eq.~\ref{eq:epnsol}:
\be
|\nabla n_{DM}(\vec r,p,t)| = \left[\sum_{i=1}^3\left(\Gamma 
\int_{t_i }^t dt_0 \, \left|\frac{\dot{p}(p_0)}{\dot{p}(p)} \right| 
\frac{dN_{e^+}}{dp_0} \, 2\,\frac{x_i-x_{i,p}(t_0)}{\lambda(p,p_0)^2}\,G(\vec r,p;\vec{r}_p(t_0), p_0)\right)^2\right]^{1/2} \;,
\label{eq:anis2}
\ee
with $p_0$ obtained from $\Delta \tau=t-t_0$ and $x_i=x,y,z$ for $i=1,2,3$, respectively.

\begin{figure}[t]
   \centering
   \includegraphics[width=7.cm]{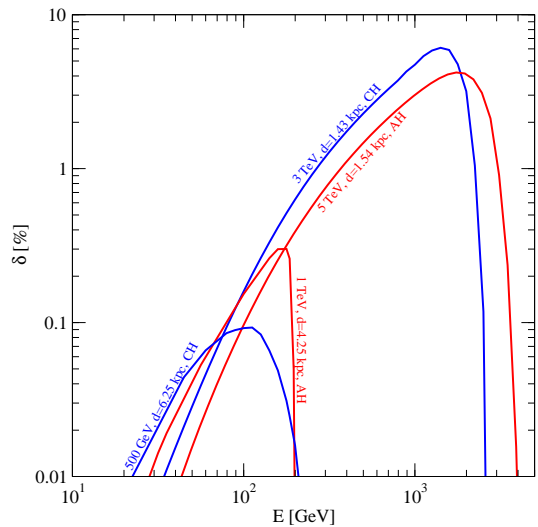}
   \caption{Dipole anisotropy of the electron plus positron spectrum as a function of the energy for the four sample DM scenarios in Table~\ref{tab:fit}.}
\label{fig:anis}
\end{figure} 

We plot in Fig.~\ref{fig:anis}, the degree of dipole anisotropy for the benchmark DM scenarios in Table~\ref{tab:fit}.
Comparing models 3 and 4, describing sources at analogous distance and with analogous local spectrum, one can note that, as intuitive, an approaching source would induce a higher anisotropy than a moving-away source (both on a straight-line trajectory). On the other hand, for sources at moderate distance the mismatch is quite small and the picture is similar to the stationary case.
The degree of anisotropy is strongly and inversely correlated to the diffusion length (see Eq.~\ref{eq:anis2}).
Although the benchmark DM models 1 and 2 induce a very similar local spectrum (see Fig.~\ref{fig:pamela1}) and describe substructures moving away from us on a vertical trajectory at the same velocity, in the second case the electrons and positrons undergo a greater diffusion before reaching us, and the anisotropy is suppressed with respect to the first case.
Note that the maximum of the curves in Fig.~\ref{fig:anis} occur at larger energy than the peaks in Figs.~\ref{fig:pamela1} and \ref{fig:pamela1b}, being the diffusion and the related wash-out of anisotropies less effective at high energy. 
The degree of dipole anisotropy for the benchmark models in Fig.~\ref{fig:anis} can be observable by FERMI in few years of data taking. On the other hand, the predictions for the anisotropy associated to a DM substructure suffer of many theoretical uncertainties, as in the case of the spectrum, and, depending on the trajectory, the degree of anisotropy can be suppressed or slightly enhanced with respect to an analogous (i.e., at analogous distance from the observer) stationary point-like case.

\section{Conclusions}
\label{sec:concl}

We have discussed the contribution to the local antimatter fluxes due to individual WIMP
DM substructures, accounting for the substructure proper motion. 
We have derived analytic solutions to the propagation equation as appropriate for a 
time-dependent positron and antiproton primary sources, identifying the relevant quantities 
to discriminate whether proper motion effects are relevant or not. 
We found that, for both positrons and antiprotons, 
the static limit is a fair approximation only in the case of high energy particles 
and nearby sources, while it fails in all other situations. The discussion has involved
a few benchmark DM candidates and and a few sample orbits for the substructure.
 
As an application of the general discussion, we focused first on WIMPs annihilating into leptons only, 
and we derived sample fits of the PAMELA positron excess and FERMI all-electron data.
The fits have been obtained starting from some arbitrary values of the WIMP mass, 
demonstrating that, for a single non-static DM point-source, it is no longer
true that one can extract from the data, in a unique way, as extensively done in recent analyses, 
model independent particle physics observables, such as the DM mass, the annihilation yield and 
the pair annihilation cross section. 
Indeed, the threshold of the local electron/positron spectrum cannot be used as a sure indicator of the DM particle mass, being possibly set by the distance of the DM substructure. The relation between the shapes of the injection spectrum and of the propagated spectrum is not straightforward for a non-static DM substructure, which behaves as a transient source.
Moreover, the annihilation rate depends on the product of the pair annihilation cross section times the unknown annihilation volume of the substructure, rather than on $\sigma v$ only.

The PAMELA positron excess can be explained in such a scenario, provided that a large annihilation rate in the substructure is considered. This requires either a large annihilation volume (much larger than typical values predicted for DM substructures in N-body simulations, but viable for scenarios accounting for the adiabatic formation of an intermediate mass black hole in the substructure) and/or a pair annihilation cross section significantly enhanced with respect to the reference thermal value. 

We have then used PAMELA and FERMI 
electron/positron data to derive, under different configurations, limits to the total annihilation 
rate in the DM source. These limits have been compared to the bounds extracted from the $\gamma$-ray luminosities associated to the same DM models. The general 
trend is that the $\gamma$-ray signal induced by a DM point-source, giving a sizable contribution to the local positron flux, is below the level of the brightest unidentified EGRET sources, but well above the FERMI $\gamma$-ray telescope sensitivity.

We have also discussed WIMP models giving raise to an antiproton yield through pair annihilations in the substructure, 
as in the case with $W^+W^-$ as final state of annihilation. 
Analogously to the electron/positron case, single DM sources inducing a significant contribution 
to the local antiproton flux can be detected by FERMI in $\gamma$-rays 
(on the other hand, as for the smooth DM halo component, 
a positron/electron flux at the level of PAMELA and FERMI data can be hardly obtained in such WIMP models 
without violating antiproton limits). 

Finally, we sketched two further features in connection to 
positron/electron emission from a DM point-source, namely the spectral and angular shape of
$\gamma$-ray flux induced by inverse Compton emission on starlight, and the dipole anisotropy 
in the electron/positron spectrum. We have shown that, potentially, both of them could be 
used to test the scenario proposed here.

\section*{Acknowledgements}
We would like to thank P.~D.~Serpico for useful discussions. 
M.R. acknowledges funding by, and the facilities of, the Centre for High Performance Computing, Cape Town. P.U. is partially supported by the European Community's Human Potential Programme under contracts  MRTN-CT-2006-035863.

\appendix

\section{Solution to the propagation equation for positrons}
\label{app:pos}

We present here a solution to Eq.~{\ref{eq:prop_n}} which is valid for a generic non-stationary
primary source of electrons and positrons in a disc-like galaxy and with the following simplifying
assumptions: We consider a model with diffusion coefficient and energy loss term that are a
function of momentum only, while they are assumed to be spatially constant within the cylindrical 
propagation volume; we impose free
escape at the vertical boundary of the diffusion region (i.e. that the equilibrium number density
vanishes at $z=\pm h_h$) and neglect the radial boundary (i.e. in the limit of radial size of the
diffusion region $R_h$ being much larger than $h_h$). Reacceleration and convection are 
also neglected. In Cartesian coordinates Eq.~(\ref{eq:prop_n}) becomes:
\be
{\partial n (\vec r,p,t) \over \partial t} =  Q(\vec r, p, t) + D(p) \Delta n 
                                                                    - {\partial\over\partial p} \left(\dot{p} \,n \right)
\label{eq:eplus_point}
\ee
The solution described here is a generalization of the one derived in Ref~\cite{Baltz:2004bb} to
the case of  diffusion coefficient $D$ and momentum gain/loss term $\dot{p}$ which are generic 
function of the particle momentum $p$; the notation we use is hence analogous to that 
of ~\cite{Baltz:2004bb}. We start by introducing $N \equiv -\dot{p} n$,  and solve the equation for 
$N$ rather than for $n$, using the Green function method and neglecting at first the vertical 
boundary condition:
\be
-\frac{1}{\dot{p}} {\partial G \over \partial t} + \frac{D}{\dot{p}} \Delta G - {\partial G \over\partial p}
=  \delta(\vec r - \vec r_0) \delta(t-t_0) \delta(p-p_0)\,.
\label{eq:eplus_green1}
\ee
Going to the Fourier transforms in the variables $\vec r$ and $t$, we find:
\be
\left[-\frac{1}{\dot{p}} (-i \omega+D |\vec k|^2)- {\partial \over\partial p}\right] \bar{G} =
 e^{i\vec k\cdot \vec r_0+i\omega t_0}  \delta(p-p_0)\,.
\label{eq:eplus_green2}
\ee
The solution of this equation is easily derived replacing the electron/positron momentum $p$ 
with a double change of variables:
\be
\tau =-\int_p^{p_{\rm max}} \frac{d\tilde{p}}{\dot{p}(\tilde{p})}
\quad \quad {\rm and} \quad \quad
\lambda^2 = 4\,  \int_{\tau_0}^\tau d\tilde{\tau} \, D(\tilde{\tau}) 
=  - 4\,  \int_p^{p_0}  {d\tilde{p}}  \, \frac{D(\tilde{p})}{{\dot{p}(\tilde{p})}}\;,
\ee
and considering first the case $p \ne p_0$. The physically relevant solution vanishes for
$p > p_0$ since we will not include energy gain effects, but only energy losses (i.e., the 
term $\dot{p}(p)$ in the equations above is negative); we find:
\be
\bar{G} = \bar{G}_0 \, e^{\left[i\,\omega \Delta \tau - |\vec k|^2 \lambda^2/4\right]}\,,
\ee
where we defined $\Delta \tau = \tau(p)-\tau(p_0)$ and the constant $\bar{G}_0$ is obtained 
integrating Eq.~(\ref{eq:eplus_green2}) around $p=p_0$:
\be
- \left. \hat{G} \right|^{p_0+}_{p_0-} = \bar{G}_0 =  e^{i\vec k\cdot \vec r_0+i\omega t_0} \,.
\ee
Taking now the inverse Fourier transform, we find the Green function:
\be
G_{free} (\vec r,t,p;\vec{r}_0,t_0,p_0)= 
\frac{\exp\left[-\frac{|\vec r - \vec r_0|^2}{\lambda^2}\right]}{\pi^{3/2} \lambda^3} 
\; \delta\left((t-t_0)-\Delta \tau\right) \;,
\ee
where it is now manifest to recognize $\lambda$ as a diffusion length scale and 
$\Delta \tau$ as an energy loss timescale; the subscript "free" was added to the Green function
since we have not included the vertical boundary condition yet. 
This is implemented with the image charge method, i.e. having defined $z_n\equiv (-1)^nz+2nh_h$, 
and having factorized the Green function as 
$G_{zb} = G_{zb-ti} \cdot  \delta\left((t-t_0)-\Delta \tau \right)$,
$G_{zb-ti}$ is given by:
\be
G_{zb-ti} (\vec r,p;\vec{r}_0,p_0)= 
\frac{\exp\left[-\frac{(x - x_0)^2+(y - y_0)^2}{\lambda^2}\right]}{\pi^{3/2} \lambda^3} \; 
\sum_{n=-\infty}^{+\infty} (-1)^n\, \exp\left[-\frac{(z_n - z_0)^2}{\lambda^2}\right] \,;
\label{eq:epgzbti}
\ee
it is easy to verify that $G_{zb}(x,y,z=\pm h_h,t;\vec{r}_0,t_0)=0$. The density per unit energy is then:
\be
n(\vec r,p,t) = -\frac{1}{\dot{p}}\,\int_{-\infty}^{+\infty}dx_0 \int_{-\infty}^{+\infty}dy_0 \int_{-h_h}^{+h_h} dz_0 \int_{-\infty}^t dt_0 \int_p^{p_{max}} dp_0\; G_{zb}(\vec r,t,p;\vec {r}_0,t_0,p_0) \,Q(\vec{r}_0, t_0,p_0)\,.  
\ee
In case the source $Q$ is point-like, such as the point-like dark matter substructure in 
Eq.~(\ref{eq:pointsource}), entering the diffusion region at the time $t_i$, the density 
per unit energy becomes the expression in Eq.~(\ref{eq:epnsol}), or equivalently takes
the form (which is the one we use for numerical calculations):
\be
n(\vec r,p,t) = \Gamma 
\int_{t_i }^t dt_0 \, \left|\frac{\dot{p}(p_0)}{\dot{p}(p)} \right| 
\frac{dN_{e^+}}{dp_0} \, G_{zb-ti} (\vec r,p;\vec{r}_p(t_0), p_0) \;,
\ee
where $p_0$ is obtained from the numerical inversion of the relation $\tau(p)-\tau(p_0) = t -t_0$.
Finally, we just remark that in case of a stationary diffuse electron/positron source, $n(\vec r,p)$ 
reduces to the familiar expression derived in the literature~\cite{Baltz:1998xv} (again generalized 
to $D$ and $\dot{p}$ which are arbitrary functions of $p$):
\be
n(\vec r,p) = -\frac{1}{\dot{p}(p)}\,\int_{-\infty}^{+\infty}dx_0 \int_{-\infty}^{+\infty}dy_0 \int_{-h_h}^{+h_h} dz_0 
\int_p^{p_{max}} dp_0\; G_{zb-ti} (\vec r,p;\vec {r}_0,p_0) \,Q(\vec{r}_0,p_0)\;,
\ee
i.e. for a stationary axisymmetric source (i.e. for $Q(\vec{r}_0,p_0)=Q(R_0,z_0, p_0)$
with $R_0$ the radial variable in a cylindrically symmetric system): 
\be
n(\vec r,p) = \int_p^{p_{max}} \frac{dp_0 }{\left|  \dot{p}(p) \right| } 
\int_{0}^{R_h} dR_0 \;  \tilde{I}_0\left(\frac{2\,R R_0}{\lambda^2}\right) 
\frac{2\,R_0\,\exp\left[-\frac{(R - R_0)^2}{\lambda^2}\right]}{\sqrt{\pi} \lambda^3} \; 
\int_{-h_h}^{+h_h} dz_0 
\sum_{n=-\infty}^{+\infty} \exp\left[-\frac{(z_n - z_0)^2}{\lambda^2}\right] \,Q(R_0,z_0,p_0)\,,
\ee
where we introduced the function $\tilde{I}_0(x)=\exp(-x)\,I_0(x)$, with $I_0(x)$ the 
modified Bessel function of first kind.

\section{Solution to the propagation equation for antiprotons}
\label{app:pbar}

Analogously to the positron case, we sketch here a solution to Eq.~{\ref{eq:prop_n}} which can 
be applied to a generic non-stationary primary source of antiprotons. We assume again that the
propagation region is a cylinder and that the diffusion coefficient is spatially constant within it; 
we impose free escape as vertical boundary condition, while neglecting the boundaries in the 
radial direction. We still do not include reacceleration and convection, as well as energy losses give a
negligible effect for antiprotons. We include instead the antiproton absorption effect, assuming it
is mainly due to a thin gas disk, with gas target density constant in space and localized at $z=0$. 
In Cartesian coordinates Eq.~(\ref{eq:prop_n}) becomes:
\be
{\partial n (\vec r,E,t) \over \partial t} =  Q(\vec r, E, t) + D(E) \Delta n- P(E) \delta(z) n
\label{eq:pbar_point}
\ee
with the term accounting for $\bar{p}$ losses due to collisions with the interstellar medium being
$P(E) = 2 h_g n_H v(E) \sigma_p(E)$, where $h_g$ is half of the gas disc height, and the gas disk is 
assumed to be made of hydrogen with constant density $n_H$. The solution sketched here is 
analogous to the one presented in Ref.~\cite{Thoudam:2008tm} in a different context. It is derived through
the Green function method, neglecting at first the vertical boundary condition:
\be
D(E) \left(\frac{\partial^2 G}{\partial x^2}
+\frac{\partial^2G}{\partial y^2}+\frac{\partial^2G}{\partial z^2}\right) - P(E) \delta(z) G
- {\partial G \over \partial t} =  -  \delta(\vec r - \vec r_0) \delta(t-t_0)
\label{eq:pbar_green1}
\ee
The following steps are to take the Fourier transforms in the $x$ and $y$ directions, i.e.
$\bar{G} = \int_{-\infty}^{+\infty} dx\int_{-\infty}^{+\infty} dy \,G e^{ik_xx+ik_yy}$:
\be
D(E) \frac{\partial^2 \bar{G}}{\partial z^2} - D(E) K^2 \bar{G} - P(E) \delta(z) \bar{G}
- {\partial \bar{G} \over \partial t} =  -   e^{ik_xx_0+ik_yy_0} \delta(z-z_0) \delta(t-t_0)
\label{eq:pbar_green2}
\ee
with $K^2 = k_x^2+ k_y^2$, and to make a Laplace transformation in the $t$ variable, i.e.
$\hat{G} = \int_0^\infty dt\,\bar{G}e^{-st}$:
\be
D \frac{\partial^2 \hat{G}}{\partial z^2} - (D K^2 + s) \hat{G} 
- P(E) \delta(z) \hat{G} =  -  e^{ik_xx_0+ik_yy_0-st_0}  \delta(z-z_0)
\label{eq:pbar_green3}
\ee
Away from $z=0$ and $z=z_0$, the equation takes the form:
\be
\frac{\partial^2 \hat{G}}{\partial z^2} - \omega^2 \hat{G} =0
\ee
with $\omega= \sqrt{K^2 + s/D}$; the three solutions in the three distinct regions (i.e. the 
region between $0$ and $z_0$,  and those below and above it, setting the appropriate 
boundary conditions at $z=\pm \infty$) are matched to find $\hat{G}$ imposing its continuity
and that (as one derives integrating Eq.~(\ref{eq:pbar_green3}) around $z=z_0$ and $z=0$):
\be
D \left.\frac{\partial \hat{G}}{\partial z}\right|^{z_0+}_{z_0-} 
+  e^{ik_xx_0+ik_yy_0-st_0} =0 \quad \quad  {\rm and} \quad \quad
D \left.\frac{\partial \hat{G}}{\partial z}\right|^{0+}_{0-} 
- P(E) \hat{G}(z=0) = 0 \;.
\ee
One finds $\hat{G} = \hat{g} \cdot e^{ik_xx_0+ik_yy_0-st_0}/D$, where
\begin{eqnarray}
\hat{g} &=& \frac{\exp\left[-\omega(\left| z \right|+\left| z_0 \right|)\right]}{P/D+2\omega} 
                     \frac{\left| sign(z)-sign(z_0)\right|}{2} \nonumber  \\
            &  & + \left\{\frac{\exp\left[-\omega\left|\left| z \right|-\left| z_0 \right|\right|\right]}{2\omega}
                                -\frac{P/D\;\exp\left[-\omega\left(\left| z \right|+\left| z_0 \right|\right)\right]}
                                         {2\omega\,(P/D+2\omega)}
                       \right\} 
                       \frac{\left| sign(z)+sign(z_0)\right|}{2}\,.
\label{eq:prop_f2}
\end{eqnarray}
This expression can be inverted analytically to recover $G$. One can take first the inverse
Laplace transform and find: 
\be
\bar{G} = e^{ik_xx_0+ik_yy_0-(k_x^2+k_y^2) \lambda_p^2/4}\; g(z,t;z_0,t_0) 
\ee
with:
\be
g(z,t;z_0,t_0) \equiv \frac{1}{\sqrt{\pi} \lambda_p}\exp\left[-\frac{(z - z_0)^2}{\lambda_p^2}\right] -
               \frac{1}{L} \exp\left[2\frac{\left| z \right|+\left| z_0 \right|}{L}+\frac{\lambda_p^2}{L^2}\right] \,
               \mathrm{erfc}\left(\frac{\lambda_p}{L}+ \frac{\left| z \right|+\left| z_0 \right|}{\lambda_p}
               \right)\;,
\ee
where we defined: $\lambda_p(E,t,t_0) = 2 \sqrt{(t-t_0)\,D(E)}$ and $L(E) = 4 D(E)/P(E)$.
Finally, taking the inverse Fourier transform we find the Green function:
\be
G_{free} (\vec r,t;\vec{r}_0,t_0)= 
\frac{1}{\pi \lambda_p^2}  \exp\left[-\frac{(x - x_0)^2+(y - y_0)^2}{\lambda_p^2}\right]
\; g(z,t;z_0,t_0) 
\ee
where the subscript "free" was added since we did not include the vertical boundary condition yet. 
As for the positron case, this is implemented with the image charge method, i.e. having 
defined $z_n\equiv (-1)^nz+2nh_h$, the Green function is:
\be
G_{zb} (\vec r,t;\vec{r}_0,t_0)= \frac{1}{\pi \lambda_p^2}  \exp\left[-\frac{(x - x_0)^2+(y - y_0)^2}{\lambda_p^2}\right] \; \sum_{n=-\infty}^{+\infty} (-1)^n\, g(z_n,t;z_0,t_0)\,.
\ee
Note that the expression we found is analogous to Eq.~\ref{eq:epgzbti} for positrons; actually 
switching off antiproton annihilations (i.e. setting $P(E)=0$) the two expressions are formally 
identical, except that the definition of diffusion length here is different from the diffusion length
introduced for positrons, with time appearing explicitly in $\lambda_p$ for antiprotons rather 
then implicitly through the momentum loss as in the $\lambda$ for positrons.   
The density per unit energy is then:
\be
n(\vec r,E,t) = \int_{-\infty}^{+\infty}dx_0 \int_{-\infty}^{+\infty}dy_0 \int_{-h_h}^{+h_h} dz_0 \int_{-\infty}^t dt_0\; G_{zb}(\vec r,t,E;\vec {r}_0,t_0) \,Q(\vec{r}_0, t_0,E)\,.  
\ee

\end{document}